\documentclass[%
 aip,
 amsmath,amssymb,
preprint,%
]{revtex4-1}

\usepackage{amssymb}
\usepackage{amsmath}
\usepackage{bm}
\usepackage{color}
\usepackage{graphicx}
\usepackage{float}
\usepackage{placeins}
\usepackage{booktabs}
\usepackage{xcolor}
\usepackage{colortbl}
\usepackage{hyperref}
\usepackage{braket}
\usepackage[inline]{enumitem}
\usepackage[caption=false]{subfig}

\newcommand{\jr}[1]{{\color{black} #1}}

\begin{document}

\title{Efficient Simulation of Non-Markovian Path Integrals via Imaginary Time Evolution of an Effective Hamiltonian}

\author{Xiaoyu Yang}
\author{Limin Liu}
\author{Wencheng Zhao}
\affiliation{Key Laboratory of Theoretical and Computational Photochemistry,
Ministry of Education, College of Chemistry, Beijing Normal University, 100875
Beijing, People's Republic of China.}
\author{Jiajun Ren}
\email{jjren@bnu.edu.cn}
\author{Wei-Hai Fang}
\affiliation{Key Laboratory of Theoretical and Computational Photochemistry,
Ministry of Education, College of Chemistry, Beijing Normal University, 100875
Beijing, People's Republic of China.}
\affiliation{Institute for Advanced Study, Beijing Normal University, 100875
Beijing, People's Republic of China.}

\begin{abstract}
Accurately simulating the non-Markovian dynamics of open quantum systems remains a significant challenge.
While the recently proposed time-evolving matrix product operator (TEMPO) algorithm based on path integrals successfully circumvents the exponential scaling associated with memory length, its reliance on layer-by-layer tensor contractions and compressions leads to steep scaling with respect to the
system Hilbert space dimension.
In this work, we introduce the effective Hamiltonian-based TEMPO (EH-TEMPO) algorithm, which reformulates the calculation of the Feynman-Vernon influence functional as an imaginary time evolution governed by an effective Hamiltonian.
We demonstrate that this effective Hamiltonian admits a highly compact matrix product operator representation, enabling substantial compression with negligible loss of accuracy.
Combining a one-shot global evolution with a backward retrieval approach, EH-TEMPO significantly reduces algorithmic complexity and is naturally suited for GPU acceleration.
We benchmark the method by simulating the energy transfer dynamics in the 7-site Fenna-Matthews-Olson complex model and 4-site perylene bisimide model.
The results demonstrate that EH-TEMPO achieves numerically exact accuracy with superior efficiency, delivering speedups of up to 17.5$\times$ on GPU hardware compared to standard CPU implementations.
\end{abstract}

\maketitle

\section{INTRODUCTION}

The study of quantum dynamics in condensed-phase systems~\cite{nitzan2006chemical,weiss2012quantum} is central to understanding a wide range of physical and chemical phenomena, including carrier transport in organic semiconductors and excitation energy transfer in photosynthetic complexes. In these systems, the phonon environment affects the electron dynamics through electron-phonon interactions, inducing decoherence and energy dissipation that ultimately govern the overall functional properties.~\cite{may2023charge,biologicalsystems2,jang2018delocalized} 

Various approaches have been developed to simulate such electron-phonon coupled dynamics.~\cite{xu2026colloquium} The first category treats the composite system, which comprises both electronic and phononic degrees of freedom, as a closed quantum system by explicitly propagating the full wavefunction according to the Schr\"{o}dinger equation. Prominent examples include the time-dependent density matrix renormalization group (TD-DMRG),\cite{ma2018time, ren2022time} tree tensor network states (TTNS),~\cite{schroder2019tensor,li2024optimal} and the multilayer multi-configuration time-dependent Hartree (ML-MCTDH) approach.\cite{multilayer1,wang2015multilayer} While these methods are highly accurate and robust across broad parameter regimes, their computational cost increases rapidly with the number of bath modes, often rendering long-time simulations of realistic condensed-phase problems prohibitively expensive.
The second category focuses on the reduced dynamics of the system of interest by tracing out the phononic degrees of freedom.
This includes perturbative approaches based on quantum master equations, such as the Lindblad\cite{lindblad1,gorini1976completely} and Redfield equations,\cite{redfield1957theory,redfield1965theory} which commonly invoke the Markov approximation.
However, capturing non-Markovian memory effects, which are essential in many chemical systems, requires numerically exact methods.
The most notable among these are the hierarchical equations of motion (HEOM),\cite{HEOM1,tanimura2020numerically} non-Markovian quantum state diffusion (NMQSD),~\cite{diosi1998non,suess2014hierarchy} and the quasi-adiabatic propagator path integral (QuAPI).\cite{makri1995tensor,makri1995tensor2}

\jr{
QuAPI, originally developed by Makri and coworkers, can in
principle fully retain non-Markovian effects and handle
arbitrary bath spectral densities at both zero and finite
temperatures.~\cite{makri1992improved, makri1995tensor,
makri1995tensor2, makri1998quantum} 
While the QuAPI algorithm scales exponentially with memory time,
various techniques have been developed to significantly improve
its efficiency in specific parameter regimes, such as path
filtering,~\cite{sim2001quantum} blip decomposition,~\cite{makri2017iterative} and coarse graining.~\cite{Ovcharenko2026Scalable}
Notably, Makri and coworkers introduced the small matrix path integral (SMatPI) algorithm,~\cite{makri2020small} which offers an efficient strategy to overcome the exponential memory storage scaling of QuAPI. By recursively disentangling the influence functional, SMatPI reduces the necessary memory storage to a small set of fixed-size matrices whose dimensions are equal to the square of system size. Another significant breakthrough was achieved with the introduction of tensor network (TN) algorithms to further enhance path integral methods.~\cite{strathearn2018efficient, ye2021constructing, cygorek2022simulation} Among them, the time-evolving matrix product operator (TEMPO) algorithm proposed by Strathearn \textit{et al}.~\cite{strathearn2018efficient} combines the influence functional formalism with TN techniques, specifically matrix product states (MPS) and matrix product operators (MPO).~\cite{schollwock2011density}
It acts as a powerful compression scheme for
the exponentially large path space. By expressing the influence
functional as the contraction of a two-dimensional TN and
utilizing singular value decomposition (SVD) algorithm, TEMPO can
systematically compress the influence functional. In broad
parameter regimes, this effectively reduces the scaling with
respect to memory length from exponential to
polynomial.~\cite{strathearn2018efficient} Intrinsically, this
efficiency is  tied to the compressibility of the
influence functional; for problems characterized by strong
system-bath coupling and long memory, capturing the strong
temporal correlations may still require a rapidly growing TN
bond dimension.  
}
Subsequent developments have significantly extended the
generality and capability of TEMPO, including reformulations
within the process tensor (PT) framework for multi-time
correlation functions,~\cite{jorgensen2019exploiting} extensions
to general off-diagonal or non-commuting system-bath couplings,
\cite{richter2022enhanced,gribben2022exact,chen2024path, zhang2025time}
generalizations to equilibrium initial
states~\cite{bose2023quantum} and fermionic baths,
~\cite{ng2023real,chen2024grassmann,chen2024grassmann2} and combinations with other
powerful algorithms, such as SMatPI~\cite{makri2020small, kundu2023pathsum} and
high-dimensional tensor networks.~\cite{bose2022multisite}

Despite these successes, the standard TEMPO algorithm still faces significant challenges when applied to multi-state systems.
A key step in TEMPO is the iterative contraction and compression of the layered MPOs representing the incremental influence functional.~\cite{strathearn2018efficient} In general, the bond dimension of MPOs grows quadratically with the system's Hilbert space dimension $d$, making standard SVD-based compression prohibitively expensive for multi-state systems, \jr{with computational costs scaling as $\mathcal{O}(d^3)$ for tensor contraction and $\mathcal{O}(d^4)$ for tensor decomposition.}
\jr{
Additionally, while the analytical construction of a compact MPO for a single system-bath coupling operator is well-documented in the literature,~\cite{strathearn2018efficient} manually constructing the MPO for general system-bath interactions remains non-trivial and laborious.
}
The existing automated MPO construction algorithms~\cite{hubig2017generic,ren2020general,ccakir2025optimal} are incompatible with the influence functional form, which lacks a simple sum-of-products (SOP) structure.
In this work, we propose an effective Hamiltonian-based TEMPO (EH-TEMPO) algorithm to address the scalability of (PT-)TEMPO for multi-state systems.
Instead of the standard layer-by-layer contraction and compression scheme, we derive the influence functional via the imaginary time evolution of an effective Hamiltonian.
This reformulation offers three distinct advantages over the original (PT-)TEMPO algorithm:
\begin{enumerate*}[label=(\roman*)]
\item The effective Hamiltonian takes an SOP form, enabling the use of automated MPO construction algorithms to generate optimal MPO representations for arbitrary system-bath couplings with minimal bond dimensions.
\item Well-established time evolution algorithms, such as those based on the time-dependent variational principle (TDVP),~\cite{haegeman2016unifying,li2020numerical} can be employed to obtain the overall influence functional in a one-shot calculation with relatively large imaginary time steps, replacing the costly layer-by-layer heavy calculations.
\item With TDVP-based time evolution algorithms, the EH-TEMPO algorithm avoids decompositions of large tensors, making it highly amenable to GPU acceleration.
\end{enumerate*}
We benchmark the EH-TEMPO algorithm by simulating excitation energy transfer dynamics in the 7-site Fenna-Matthews-Olson (FMO) model \jr{and 4-site perylene
bisimide (PBI) model}.
By comparing our results with those from the PT-TEMPO algorithm, we demonstrate that EH-TEMPO provides accurate influence functionals with superior computational efficiency for multi-state systems.

The remainder of this paper is organized as follows. Section~\ref{second:methods} briefly reviews the theoretical foundations of the QuAPI and TEMPO algorithms and then details the proposed EH-TEMPO algorithm.
In Section~\ref{fourth:result}, we systematically analyze the accuracy and efficiency of the EH-TEMPO method, comparing its performance against standard benchmarks.
Finally, conclusions are presented in Section~\ref{conclusion}.

\section{METHODS}
\label{second:methods}

\subsection{The QuAPI and TEMPO Algorithms}
\begin{figure}[htbp]
\centering
\includegraphics[width=0.99\textwidth]{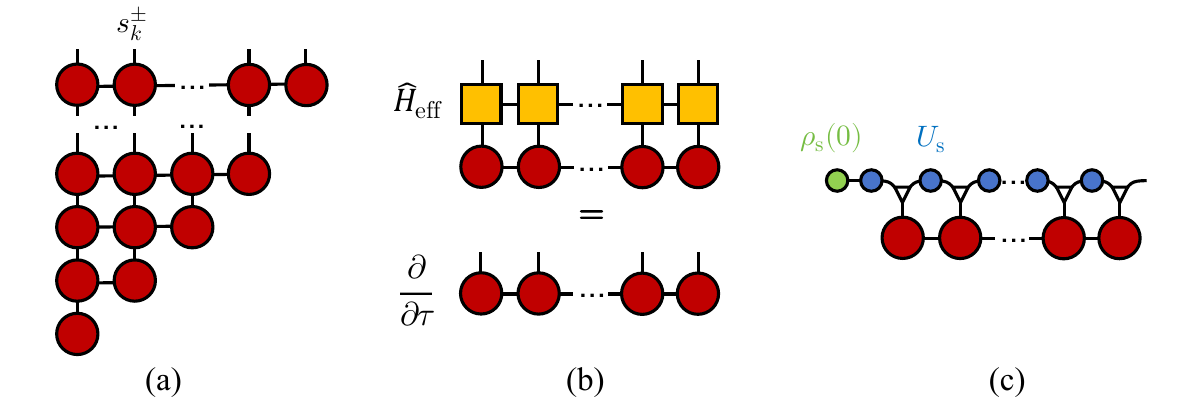}
\caption{
Tensor network diagrams for the TEMPO algorithms. (a) The triangular tensor network representation of the influence functional used in the (PT-)TEMPO algorithm.
(b) The core algorithm of EH-TEMPO, where the influence functional is generated via the imaginary time evolution governed by the effective Hamiltonian MPO ($\hat{H}_{\text{eff}}$).
(c) The final tensor contraction to obtain the reduced density matrix, combining the influence functional MPS with the system propagators $U_S$ and the initial state $\rho_S(0)$.
The white triangle represents $\delta_{ijk}$.
}
\label{fig:graph_scheme}
\end{figure}

A generic system-bath coupled problem is described by the total Hamiltonian:
\begin{align}
\hat{H} & = \hat{H}_S + \hat{H}_B + \hat{H}_{SB}, \\
\hat H_B & = \sum_i \frac{1}{2} \hat p_i^2 + \frac{1}{2} \omega_i^2 \hat q_i^2, \\
\hat H_{SB} & =  \sum_{n} \hat{S}_n \otimes \sum_i c_{ni} \hat{q}_i, \label{eq:H_SB}
\end{align}
where $\hat{H}_S$ is the system Hamiltonian, $\hat{H}_B$ is the bath Hamiltonian, and $\hat{H}_{SB}$ describes the system-bath interaction.
In this work, we restrict our consideration to diagonal system-bath coupling, for which there exists a basis $\{|s\rangle\}$ that simultaneously diagonalizes all operators $\hat{S}_n$, with eigenvalues $s$.
For notational conciseness, we consider a single coupling term in Eq.~\eqref{eq:H_SB} in the following and thus omit the summation over $n$.
The time evolution of the total density matrix is given by $\hat{\rho}(t) = e^{-i\hat{H}t}\hat{\rho}(0)e^{i\hat{H}t}$.
We assume a factorized initial condition at $t=0$, $\hat{\rho}(0) = \hat{\rho}_S(0) \otimes \hat{\rho}_B^{\text{eq}}$, where the bath is in thermal equilibrium.
The extension to correlated thermal equilibrium initial states has been discussed in Refs.~\onlinecite{shao2001iterative, shao2002iterative,bose2023quantum}.
Applying the standard Trotter splitting of the short-time
propagator $e^{-i\hat{H} \Delta t}$ to separate system and bath
contributions, the reduced density matrix $\hat{\rho}_S(t) =
\text{Tr}_B[\hat{\rho}(t)]$ at time $t = N_t \Delta t$ can be
expressed as a discretized path integral:
\begin{align}
\label{master}
\hat{\rho}_{S}(N_t\Delta t) &= \sum_{s_0^\pm, \dots, s_{N_t-1}^\pm}
\langle s_{0}^{+} |
\hat{\rho}_{S}(0) | s_{0}^{-} \rangle \,
G(s_0^\pm,  \dots, s_{N_t}^\pm) \,
F(s_1^\pm,  \dots, s_{N_t}^\pm).
\end{align}
Here, $s_0^\pm, s_1^\pm, \dots, s_{N_t}^\pm$ denote the discretized forward ($+$) and backward ($-$) system paths.
The system propagator $G$ is given by the product of short-time propagators:
\begin{equation}
\label{sys_tensor}
G(s_0^\pm,  \dots, s_{N_t}^\pm) = \prod_{k=1}^{N_t} \langle s_{k}^{+} |
e^{-i\hat{H}_0 \Delta t} | s_{k-1}^{+} \rangle \langle s_{k-1}^{-} | e^{i\hat{H}_0 \Delta t} |
s_{k}^{-} \rangle,
\end{equation}
where $\hat H_0$ is the QuAPI-type shifted system Hamiltonian along the adiabatic path.~\cite{makri1992improved}
The environmental effects are encoded in the Feynman-Vernon influence functional $F$, which accounts for non-Markovian memory effects.~\cite{feynman2000theory} 
\jr{
It should be noted that we do not employ any memory truncation in the expressions and calculations in the next section; the full non-Markovian history is retained for the entire propagation time.
}
For a Gaussian bath with linear coupling, $F$ takes the analytical form:~\cite{makri1995tensor}
\begin{align}
\begin{split}
\label{inf}
    &F\left(s_1^\pm,  \dots, s_{N_t}^\pm\right) = \prod_{k=1}^{N_t}
    \prod_{k'=1}^{k} I_{\Delta k}(s_{k'}^{\pm},s_{k}^{\pm}), \quad \Delta k = k-k', \\
    &I_{\Delta k}(s_{k'}^{\pm},s_{k}^{\pm})=\exp{\left[-(s_k^+-s_k^-)(\eta_{kk'}
    s^+_{k'}-\eta_{kk'}^*s_{k'}^-)\right]}.
\end{split}
\end{align}
The coefficients $\eta_{kk'}$ are determined by the bath spectral density and the temperature. 
\jr{
For simplicity, we present expressions based on first-order Trotter splitting ($e^{-iH\Delta t} \approx e^{-i\hat H_0\Delta t} e^{-i (\hat H-\hat H_0)\Delta t}$ with local error $\mathcal{O}(\Delta t^2)$).
The extension to second-order splitting is straightforward ($e^{-iH\Delta t} \approx e^{-i\hat H_0\Delta t/2} e^{-i (\hat H-\hat H_0)\Delta t} e^{-i\hat H_0\Delta t/2} $ with local error $\mathcal{O}(\Delta t^3)$).~\cite{makri1995tensor,makri1995tensor2}
}
Computing Eq.~\eqref{master} exactly incurs an exponential cost with respect to the memory length. 
To overcome this challenge, the TEMPO algorithm identifies the discrete path variables $s_k^{\pm}$ as the physical indices of a tensor network.
The incremental influence functional in Eq.~\eqref{eq:if_incremental} can be compactly represented as a one-dimensional diagonal MPO (after inserting Kronecker deltas with respect to $s_k^{\pm}$):
\begin{gather}
    f_k(s_1^\pm,\cdots,s_k^\pm) = \prod_{k'=1}^kI_{\Delta k}(s_{k'}^\pm,s_k^\pm), \label{eq:if_incremental}
\\
  \hat B_k = \left(\prod_{k'=1}^{k-1}\delta_{i_{k'}^+}^{s_{k'}^+} \delta_{i_{k'}^-}^{s_{k'}^-} \right) f_k = \sum_{\{a\}} B[1]^{s_1^\pm,i_1^\pm}_{a_1}  \cdots B[k-1]^{s_{k-1}^\pm,i_{k-1}^\pm}_{a_{k-2}a_{k-1}}  B[k]^{s_k^\pm}_{a_{k-1}},
\label{eq:if_incremental_mpo}
\end{gather}
Here, $a_i$ is the index linking adjacent matrices, the size of which is called the bond dimension. The overall influence functional thus exhibits a two-dimensional triangular structure, which is contracted as a sequence of layered MPOs to yield a final MPS (see Fig.~\ref{fig:graph_scheme}(a)):
\begin{align}
  F\left(s_1^\pm,  \dots, s_{N_t}^\pm \right) & = \hat B_{N_t}  \hat  B_{N_t-1}\cdots  \hat  B_1 \\
& \approx \sum_{\{a\}} A^{s_1^\pm}_{a_1} A^{s_2^\pm}_{a_1a_2} \cdots A^{s_{N_t}^\pm}_{a_{N_t-1}} .
\end{align}
In practice, after successively contracting adjacent layers, standard SVD-based compression is applied to truncate the bond dimension of MPS by discarding renormalized states with small singular values, thereby circumventing the exponential scaling of computational cost and storage with respect to $N_t$.

In the original TEMPO algorithm,\cite{strathearn2018efficient} starting from the system initial state $\rho_S(0)$, the one-step system propagator is absorbed into the pairwise influence functional with $\Delta k=1$ during the contraction.
Consequently, instead of obtaining the overall influence functional, the so-called augmented density tensor (ADT), which is the summand in Eq.~\eqref{master}, is obtained directly.
The subsequent combination with the PT framework\cite{jorgensen2019exploiting} enables the system propagator to be separated from the influence functional construction, resulting in higher accuracy for a given bond dimension in our experience.
Moreover, PT-TEMPO allows reuse of the process tensor for different system Hamiltonians and operations, and is especially well suited for computing multi-time correlation functions.
The trade-off is that the index of each local matrix must in general be doubled, $s_k \to (s_k, s_k')$, which inevitably increases the computational cost.
Fortunately, this doubling can be avoided for the diagonal system-bath interaction considered in this work, making the computational cost of PT-TEMPO comparable to that of TEMPO.
\jr{
It is important to emphasize that the process tensor encodes the exact same physical information as the foundational Feynman-Vernon influence functional. Mathematically, the process tensor is isomorphic to the time-discretized evaluation of the influence functional, reformulated as a discrete, multilinear operator to facilitate efficient tensor network operations.
}
Once the overall influence functional is obtained in MPS form, the system reduced density matrix is readily computed by contracting with the system propagators and initial state (see Fig.~\ref{fig:graph_scheme}(c)).

We now analyze the computational scaling of TEMPO and PT-TEMPO for diagonal system-bath couplings. 
Building upon the original algorithm proposed in Ref.~\onlinecite{strathearn2018efficient}, our implementation separates the forward and backward index variables into $s_1^+, s_1^-, s_2^+, s_2^-, \cdots, s_{N_t}^+, s_{N_t}^-$ rather than $s_1^\pm, s_2^\pm, \cdots, s_{N_t}^\pm$. Although this doubles the total number of sites, it significantly reduces the computational cost per site.
In the equations and figures presented in this paper, however, we group the forward and backward indices into a single site for clarity.
the computational complexity of tensor contraction is $\mathcal{O}(N_t^2 M_S^2 d^3)$, where $M_O$ is the bond dimension of the MPO in Eq.~\eqref{eq:if_incremental_mpo}, $M_S$ is the bond dimension of the MPS representing the overall influence functional (PT-TEMPO) or ADT (TEMPO), $d$ is the system Hilbert space dimension, and $N_t$ is the number of time steps.
After contraction, the scaling of tensor decomposition for canonicalization and compression is $\mathcal{O}(N_t^2 M^3_S d^4)$.~\cite{li2020numerical}
\jr{
The detailed analysis of computational complexity is presented in the Appendix~\ref{appendix:scaling}.
}
The standard TEMPO and PT-TEMPO algorithms thus exhibit steep scaling with respect to the system dimension $d$, especially in tensor decomposition, which prohibits the efficient treatment of multi-state dynamics.
This limitation motivate the development of the EH-TEMPO algorithm presented in the next section.

\subsection{The EH-TEMPO Algorithm}
\label{third:piet-eh}

The central idea of the EH-TEMPO algorithm is the reformulation of the influence functional as an imaginary time evolution  (ITE) governed by an effective Hamiltonian $\hat{H}_{\text{eff}}$, which acts on the space of path indices $|s_1^\pm \rangle \otimes |s_2^\pm\rangle \otimes \cdots\otimes |s_{N_t}^\pm\rangle$:
\begin{equation}
\label{eff}
\hat{H}_{\text{eff}}^{N_t}= \sum_{k=1}^{N_t} \sum_{k'=1}^k (\hat{s}_k^+ - \hat{s}_k^-) \left( \eta_{kk'} \hat{s}_{k'}^+ - \eta_{kk'}^* \hat{s}_{k'}^- \right).
\end{equation}
Here, the operators $\hat{s}_k^\pm$ are diagonal in the path basis, satisfying $\hat{s}_k^\pm |s_k^\pm\rangle = s_k^\pm |s_k^\pm\rangle$.
This effective Hamiltonian describes a one-dimensional lattice
of length $N_t$ along the temporal axis with long-range,
complex-valued pairwise interactions $\eta_{kk'}$ and $\eta_{kk'}
^*$.
For example, for a two-level system where $s_k^\pm$ takes values $\pm 1$, $\hat{H}_{\text{eff}}$ is analogous to an Ising chain with long-range couplings.
\jr{As recognized in early literature, this connection arises because the path integral representation of the spin-boson partition function is exactly equivalent to a classical one-dimensional Ising model with long-range couplings.~\cite{leggett1987dynamics}}
Since $\hat{H}_{\text{eff}}$ takes the form of a sum of products,
its MPO representation can be efficiently constructed using the
automated bipartite graph algorithm developed in our previous
works.\cite{ren2020general,li2024optimal} 
\jr{
The automated bipartite graph algorithm can convert a general Hamiltonian, expressed as a sum of local operator products, into a highly compact MPO without requiring manual derivation. By mapping the operator strings onto a bipartite graph structure, where nodes represent local operators and edges represent interactions, the algorithm symbolically identifies and merges shared complementary operator prefixes and suffixes across all terms. This graph-based optimization effectively eliminates redundant local operators and maximally reduces the internal MPO bond dimensions (but still exact). Consequently, it provides a highly efficient programmatic route to construct the MPO representation for complex effective Hamiltonians, directly ensuring optimal memory and computational performance during subsequent tensor network calculations.
}
To compute the state $|\Psi_{\text{IF}}\rangle = \sum_{s_1^\pm,  \dots, s_{N_t}^\pm} F(s_1^\pm,  \dots, s_{N_t}^\pm) |s_1^\pm,  \dots, s_{N_t}^\pm\rangle$, which encodes the overall influence functional from
$t_1$ to $t_{N_t}$ in its amplitudes, we start from an unnormalized initial state $|\Psi_0\rangle$ representing an equal superposition of all path configurations.
This state admits an MPS representation with $M_S=1$ (All elements in each local site are 1, $A_{a_{i-1}a_{i}}^{s_i^{\pm}}=1$):
\begin{gather}
|\Psi_0\rangle = \sum_{s_1^\pm, \dots, s_{N_t}^\pm} \left| s_1^\pm, \dots, s_{N_t}^\pm \right\rangle.\label{equal amplitude} 
\end{gather}
Then, $ |\Psi_{\text{IF}}\rangle$ can be calculated by solving
the imaginary time Schr\"{o}dinger equation from $\tau=0$ to
$\tau=1$ under the effective Hamiltonian (see Fig~\ref{fig:graph_scheme}(b)),
\jr{
\begin{gather}
\frac{\partial}{\partial \tau} |\Psi(\tau)\rangle = -\hat{H}_{\mathrm{eff}} |\Psi(\tau)\rangle, \quad |\Psi(0)\rangle = |\Psi_0\rangle, \\
 |\Psi_{\text{IF}}\rangle = |\Psi(\tau)\rangle = e^{-\tau \hat{H}_{\text{eff}}} |\Psi_0\rangle, \quad \tau=1. \label{IF_mps}
\end{gather}
In this work, we employ the standard TDVP-based projector
splitting (PS) algorithm to perform the
ITE.\cite{haegeman2016unifying,li2020numerical} 
The TDVP-PS algorithm is an effcient algorithm for integrating the time-dependent Schrödinger equation strictly within the MPS manifold, under the Dirac-Frenkel TDVP $\langle \delta \Psi | i\partial /\partial t -\hat H |\Psi \rangle = 0$. Rather than applying the evolution operator globally and subsequently truncating the bond dimension, TDVP directly projects the time evolution operator onto the tangent space of the current MPS. The core mechanism of the splitting approach lies in recognizing that the global tangent space projector can be mathematically decomposed into a sequence of local projectors acting on individual site tensors, alternating with negative projectors acting on the intervening bond matrices. By applying a Trotter decomposition to this operator, the global time evolution is transformed into a robust, DMRG-like sweeping algorithm. During a sweep, each local site tensor (the canonical center) is evolved forward in time, followed by a backward-in-time evolution of the adjacent bond matrix, which correctly shifts the canonical center to the next site while preventing the double-counting of local dynamics. In our calculations, we employ the one-site TDVP-PS algorithm, which ensures the bond dimension remains fixed throughout the evolution. This practically prevents the exponential growth in computational cost and memory storage associated with the number of time steps. Consequently, systematic convergence checks with respect to the chosen maximum bond dimension must be performed in practical applications.
}

In (PT-)TEMPO, if one regards each layer of MPO contraction as one time step, obtaining $|\Psi_{\text{IF}}^{N_t}\rangle$ requires $N_t$ steps.
By contrast, the number of ITE steps $N_\tau$ required in EH-TEMPO can be very small, typically from several to a dozen steps suffice, with $N_\tau \ll N_t$, as shown in the next section.
This substantial reduction in the number of evolution steps translates directly into computational savings.
Moreover, sophisticated adaptive time-step algorithms can be employed to further automate the ITE process, which is important for investigating unknown problems.
Once $|\Psi_{\text{IF}}^{N_t}\rangle$ is obtained, the reduced density matrix at the final time step $t_{N_t}$ is computed as in PT-TEMPO (see Fig.~\ref{fig:graph_scheme}(c)).
However, a direct application of EH-TEMPO provides only the reduced density matrix at the final time step $t_{N_t}$.
To obtain the full dynamical trajectory from $t_1$ to $t_{N_t}$, three schemes can be employed:

\begin{enumerate}
\item \textbf{Independent scheme.} Perform ITE independently for $\hat H_\textrm{eff}^1$, $\hat H_\textrm{eff}^2$, $\cdots$, $\hat H_\textrm{eff}^{N_t}$ to obtain $|\Psi_{\text{IF}}^1\rangle$, $|\Psi_{\text{IF}}^2\rangle$, $\cdots$, $|\Psi_{\text{IF}}^{N_t}\rangle$ separately.
Assuming $N_\tau$ steps are required for each evolution to $\tau=1$, this scheme requires $N_t \cdot N_\tau$ ITE steps in total.
\item \textbf{Stepwise scheme.} Decompose $\hat H_\textrm{eff}^{N_t}$ into a sum of incremental contributions $\hat h_\textrm{eff}^k$ defined in Eq.~\eqref{eq:sub_h}, and perform ITE sequentially according to $\hat h_\textrm{eff}^1$, $\hat h_\textrm{eff}^2$, $\cdots$, $\hat h_\textrm{eff}^{N_t}$, each from $\tau=0$ to $\tau=1$:
\begin{gather}
\hat h_\textrm{eff}^k = \sum_{k'=1}^k (\hat{s}_k^+ - \hat{s}_k^-) \left( \eta_{kk'} \hat{s}_{k'}^+ - \eta_{kk'}^* \hat{s}_{k'}^- \right).
\label{eq:sub_h}
\end{gather}
This approach closely parallels (PT-)TEMPO, with the key difference that the incremental influence functional is exactly and explicitly constructed in (PT-)TEMPO, whereas it is implicitly obtained through ITE in EH-TEMPO.
Like the independent scheme, this requires $N_t \cdot N_\tau$ ITE steps in total.
\item \textbf{Backward retrieval scheme.} All intermediate states $|\Psi_{\text{IF}}^{n<N_t}\rangle$ are successively retrieved from $|\Psi_{\text{IF}}^{N_t}\rangle$ by exploiting a property of the influence functional: when $s_N^+ = s_N^-$, the incremental influence functional $f_N$ equals unity.
This allows the influence functional to be retrieved backward:
\begin{equation}
|\Psi_{\text{IF}}^{N_t}\rangle  \xrightarrow{s_{N_t}^+ = s_{N_t}^-} |\Psi_{\text{IF}}^{N_t-1}\rangle \xrightarrow{s_{N_t-1}^+ = s_{N_t-1}^-} \cdots \xrightarrow{s_{2}^+ = s_{2}^-}  |\Psi_{\text{IF}}^{1}\rangle.
\end{equation}
Combined with this backward retrieval, only a one-shot evolution is needed, requiring just $N_\tau$ ITE steps, which is significantly fewer than the preceding two schemes.
The retrieval itself incurs negligible additional cost, involving only a single MPS sweep.
However, since only $|\Psi_{\text{IF}}^{N_t}\rangle$ is targeted in the one-shot calculation, the retrieval may introduce small errors.
\end{enumerate}

The computational cost of EH-TEMPO is dominated by the ITE.
Among the three schemes, the backward retrieval scheme is the most efficient and will be adopted in the numerical calculations presented in Section~\ref{fourth:result}.
Using the TDVP projector splitting method, the tensor contraction scales as $\mathcal{O}(l  N_t N_\tau (M_S^2 M_O^2 d + M_S^3 M_O d))$, and the tensor decomposition scales as $\mathcal{O}(N_t N_\tau  M_S^3 d)$. 
\jr{
The detailed analysis of computational complexity is presented in the Appendix~\ref{appendix:scaling}.
$l$ is the number of tensor contraction demanded by the initial value problem (IVP) solver in updating each local site (typically $l \sim 10$). 
}
\jr{
Compared to (PT-)TEMPO, the computational scaling of tensor decomposition in EH-TEMPO is significantly reduced with respect to both the system size $d$ and the number of time steps $N_t$. For tensor contraction, the relative efficiency depends on $M_O$, the bond dimension of the effective Hamiltonian, which is determined by the system-bath coupling operator. For a single bath with one system-bath coupling term, such as in the Caldeira-Leggett model, the exact MPO has $M_O = \mathcal{O}(N_t)$, independent of $d$. For the Frenkel-Holstein model, where each local two-level system couples to an independent bath, the exact MPO scales as $M_O = \mathcal{O}(N_t d)$. Furthermore, as demonstrated in Section~\ref{fourth:result} and Appendix~\ref{appendix:scaling}, we found that the effective Hamiltonian MPO can be highly compressed with minimal error, regardless of whether the bath possesses a structureless or structured spectral density. After compression, $M_O$ is substantially smaller than its exact (uncompressed) value. Most importantly, this compressed bond dimension becomes independent of $N_t$ (see Appendix~\ref{appendix:scaling}). This intrinsic compressibility, which has also been observed in Ref.~\onlinecite{guo2024efficient}, yields significant computational savings.
}
In addition, since EH-TEMPO avoids the decomposition of large tensors and the dominant operations are large tensor contractions, it is highly amenable to GPU parallelization, as demonstrated in Ref.~\onlinecite{li2020numerical}.
This leads to the significant speedups reported in Section~\ref{fourth:result}.

\jr{
Before closing this section, we briefly contrast EH-TEMPO with recent, closely related methods based on the TEMPO framework.~\cite{zhang2025time, guo2024efficient} In Ref.~\onlinecite{zhang2025time}, Shi and coworkers derived a partial differential equation to compute the incremental influence functional. While technically similar to the stepwise scheme introduced above, our approach differs by adopting an effective Hamiltonian interpretation to calculate the overall influence functional in a  one-shot calculation. This effectively reduces the computational scaling from $\mathcal{O}(N_t^2)$ to $\mathcal{O}(N_\tau N_t)$, where $N_\tau \ll N_t$. In Ref.~\onlinecite{guo2024efficient}, Guo and coworkers employ Prony fitting to approximate $\eta_{\Delta k}$ and construct the MPO for the exponent of fermionic influence functional by exploiting the time-translational invariance of $\eta_{\Delta k}$. The subsequent time evolution is then obtained using the W\textsuperscript{I/II} and exponential thermal tensor renormalization group (XTRG) algorithms, which is also a one-shot calculation. Besides the differences in the time-evolution algorithms, EH-TEMPO utilizes a different MPO construction strategy: it maps the effective Hamiltonian to an exact MPO and subsequently compresses it using SVD without assuming a specific function form, which is inherently more numerically robust and stable than Prony fitting.
}

\section{RESULTS AND DISCUSSION}
\label{fourth:result}

To evaluate the performance of the EH-TEMPO algorithm, we apply it to the widely used testbed, the excitation energy transfer dynamics in molecular aggregates described by the Frenkel-Holstein model. The Hamiltonian is:
\begin{align}
\label{eq-hs}
\hat{H}_{S} & = \sum_{i} \varepsilon_i \hat{a}_i^\dagger \hat{a}_i + \sum_{i \neq j} J_{ij} \hat{a}_i^\dagger \hat{a}_j, \\
\hat{H}_{B} + \hat{H}_{SB} & = \sum_{i, \xi} \omega_{\xi} \hat{b}_{i\xi}^\dagger \hat{b}_{i\xi} + \sum_{i, \xi} g_{\xi} \omega_{\xi} \hat{a}_i^\dagger \hat{a}_i (\hat{b}_{i\xi}^\dagger + \hat{b}_{i\xi}),\label{eq-hb}
\end{align}
where $\varepsilon_i$ is the site energy of molecule $i$, and $J_{ij}$ is the excitonic coupling between molecules $i$ and $j$. Each molecule has an independent phonon bath with frequency $w_\xi$ and dimensionless electron-phonon coupling constant  $g_{\xi}$. 

For the Frenkel-Holstein model, the influence functional generalizes Eq.~(\ref{inf}) by taking the product over all site indices:
\begin{align}
\label{eq:IF_FMO}
    F(s_1^\pm,  \dots, s_{N_t}^\pm) = \prod_{i=1}^{d} \exp\left\{-  \sum_{k=1}^{N_t} \sum_{k'=1}^{k} \left( {n}_k^{i+} - {n}_k^{i-} \right) \left( \eta_{kk'} {n}_{k'}^{i+} - \eta_{kk'}^* {n}_{k'}^{i-} \right)\right\},
\end{align}
where $|s_k^\pm\rangle$ takes the local electronic state $|i\rangle$ with $i=1,\cdots,d$ ($d$ is the number of molecules), and $n_k^{i\pm}$ is the occupation number of site $i$ at step $k$.
The corresponding effective Hamiltonian is
\begin{gather}
\label{eq:Heff_FMO}
    \hat H_\textrm{eff} = \sum_{i=1}^{d} \sum_{k=1}^{N_t} \sum_{k'=1}^{k} \left( \hat{n}_k^{i+} - \hat{n}_k^{i-} \right) \left( \eta_{kk'} \hat{n}_{k'}^{i+} - \eta_{kk'}^* \hat{n}_{k'}^{i-} \right),
\end{gather}
where $\hat{n}_k^{i\pm}$ is the occupation number operator satisfying $\hat{n}_k^{i\pm} |s_k^\pm\rangle = \delta_{is} |
s_k^\pm\rangle$.

\jr{We benchmarked our approach using two model systems: the 7-site Fenna-Matthews-Olson complex with a Debye spectral density, and a 4-site perylene bisimide aggregate coupled to discrete bath modes. All calculations were performed using the Renormalizer package.}

\subsection{FMO model}

\begin{figure}[htbp]
\centering
\includegraphics[width=0.6 \columnwidth]{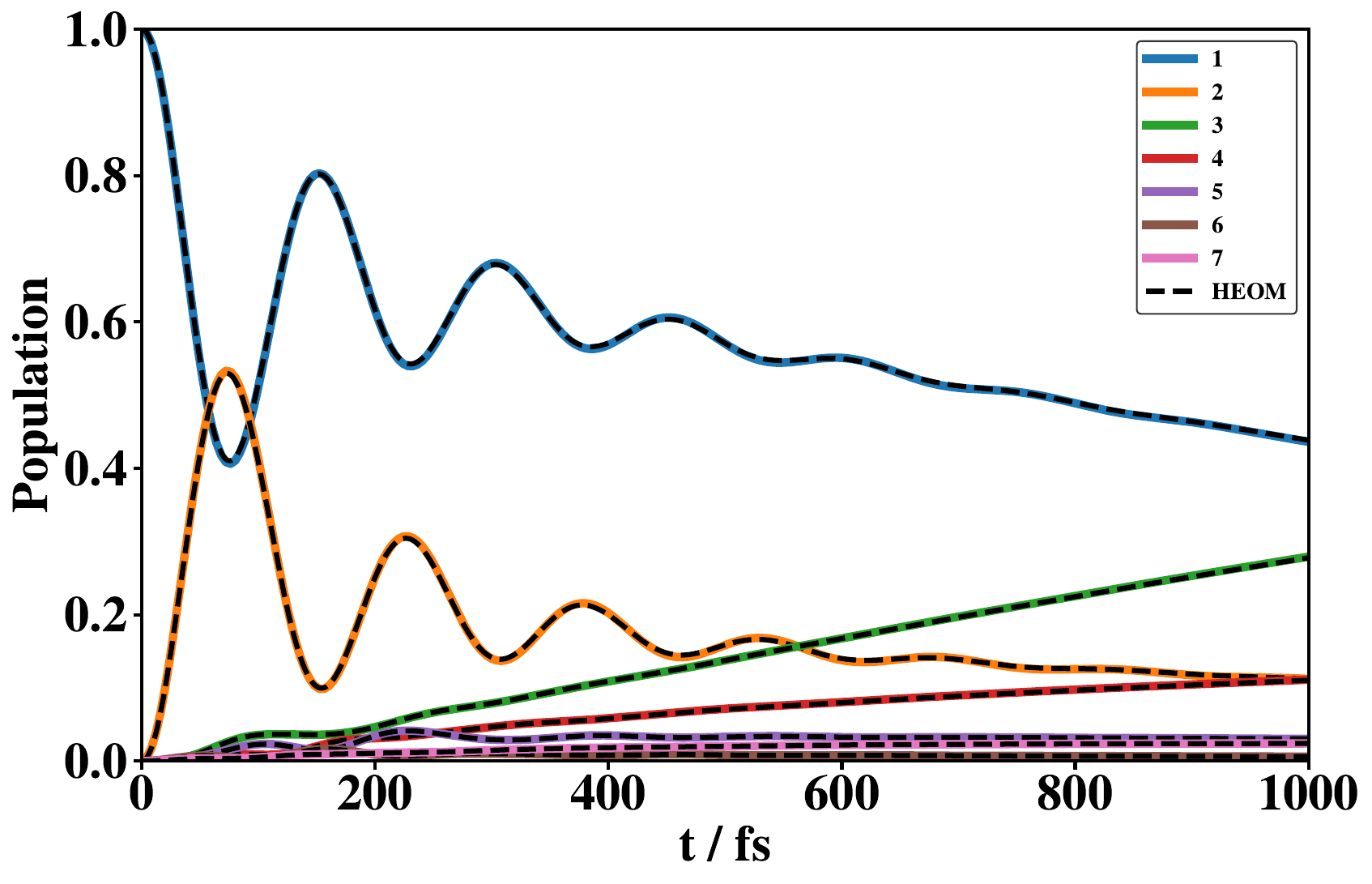}
\caption{
Exciton population dynamics of the 7-site FMO complex at $T=77\,\text{K}$. The solid colored lines represent the results obtained via the EH-TEMPO algorithm (bond dimension $M_S=128$), while the dashed black lines denote the numerically exact HEOM reference. The two datasets are in excellent agreement, with the EH-TEMPO results accurately reproducing the detailed coherent oscillations.}
\label{piiet_m100}
\end{figure}

For the FMO model, the parameters for $\hat H_S$ are from Refs.~\onlinecite{fmo_sys,fmo_sys2}.
The bath is characterized by a Debye spectral density,
$J(\omega) = 2 \lambda \frac{\omega \omega_c}{\omega^2 + \omega_c^2}$,
where $\lambda$ is the reorganization energy and $\omega_c$ is the cutoff frequency. We benchmarked EH-TEMPO against the numerically exact HEOM method and the PT-TEMPO algorithm. For the Debye spectral density, the HEOM method serves as an ideal reference.
The accuracy of the population dynamics is quantified by the time-averaged cumulative deviation:
\begin{equation}
\label{eqqor}
\mathcal{E}(\Delta t N_t) = \frac{1}{7 N_t} \sum_{i=1}^{7} \sum_{k=1}^{N_t} \left|
P^{(i)}(k) - P_{\text{ref}}^{(i)}(k) \right| ,
\end{equation}
where $P^{(i)}$ denotes the population of site $i$.
We first verify the correctness of the EH-TEMPO algorithm. Fig.~\ref{piiet_m100} displays the population dynamics at $T = 77\,\text{K}$ with $\lambda = 35\,\text{cm}^{-1}$ and $\omega_c = 1/50\,\text{fs}^{-1}$.
The EH-TEMPO results employ a real-time step size $\Delta t = 4\,\textrm{fs}$, an adaptive imaginary time step (with relative tolerence $10^{-5}$), and a bond dimension $M_S=128$.
The results show excellent agreement with the HEOM reference, capturing all oscillatory features of the coherent energy transfer dynamics.

\begin{figure}[htbp]
\centering
\includegraphics[width=0.6 \columnwidth]{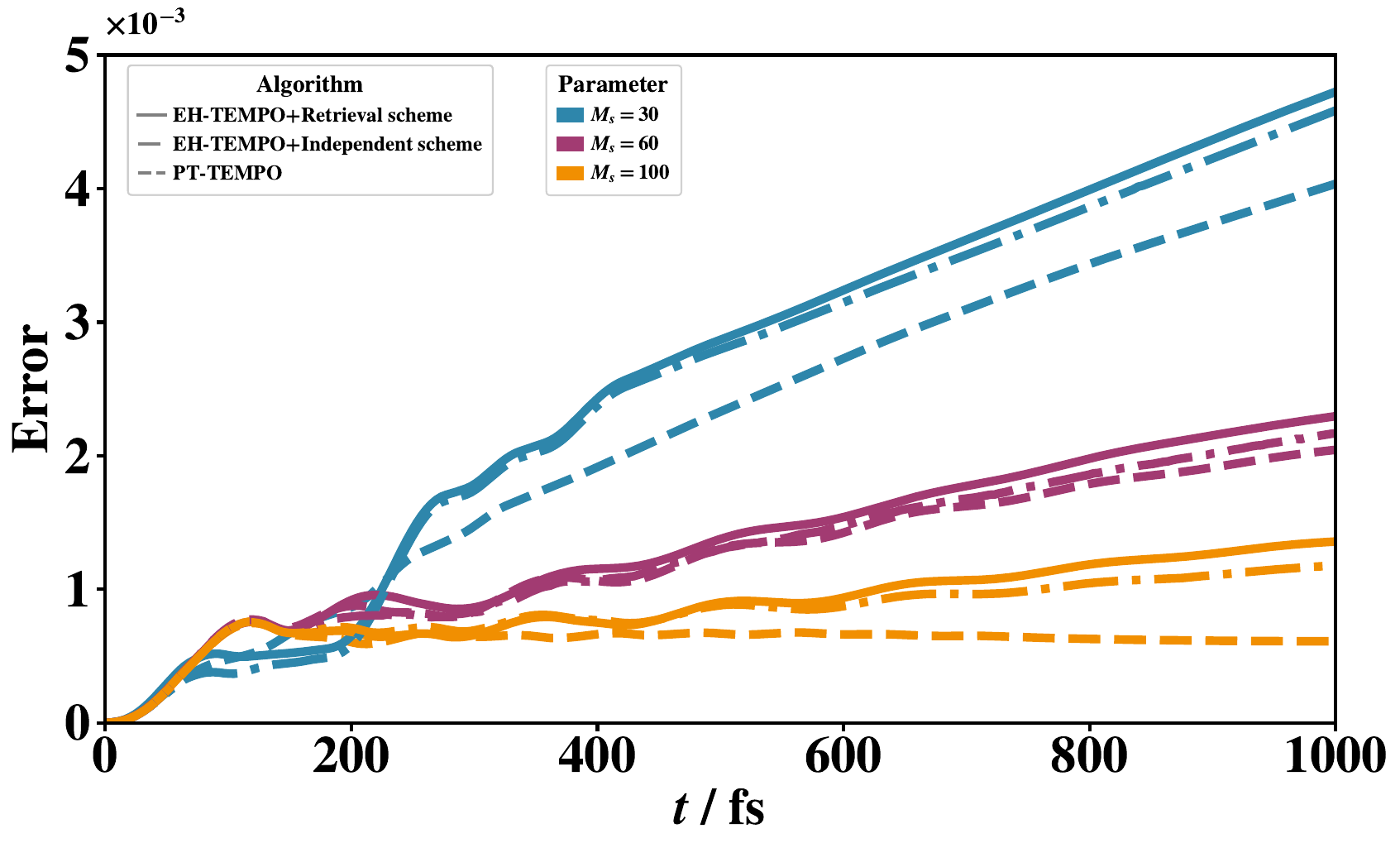}
\caption{
Time evolution of the cumulative error $\mathcal{E}(t)$ for the 7-site FMO complex. The performance of EH-TEMPO is compared against PT-TEMPO across three bond dimensions ($M_S=30, 60, 100$), represented by different colors. The line styles distinguish the algorithmic schemes: solid lines for EH-TEMPO with the backward retrieval scheme, dash-dotted lines for the independent scheme, and dashed lines for PT-TEMPO. 
}
\label{m_tempo_vs_imag}
\end{figure}

In Fig.~\ref{m_tempo_vs_imag}, we compare the error accumulation $\mathcal{E}(t)$ of EH-TEMPO against PT-TEMPO.
Two schemes are tested for EH-TEMPO: the backward retrieval scheme and the independent scheme.
Both EH-TEMPO schemes maintain errors on the order of $10^{-3}$, comparable to PT-TEMPO at the same bond dimension $M_S$.
Crucially, the backward retrieval scheme achieves accuracy similar to the much more expensive independent scheme, confirming that with a proper bond dimension the efficient backward retrieval introduces negligible additional  error.
\FloatBarrier

\begin{figure}[htbp]
\centering
\includegraphics[width=0.99 \columnwidth]{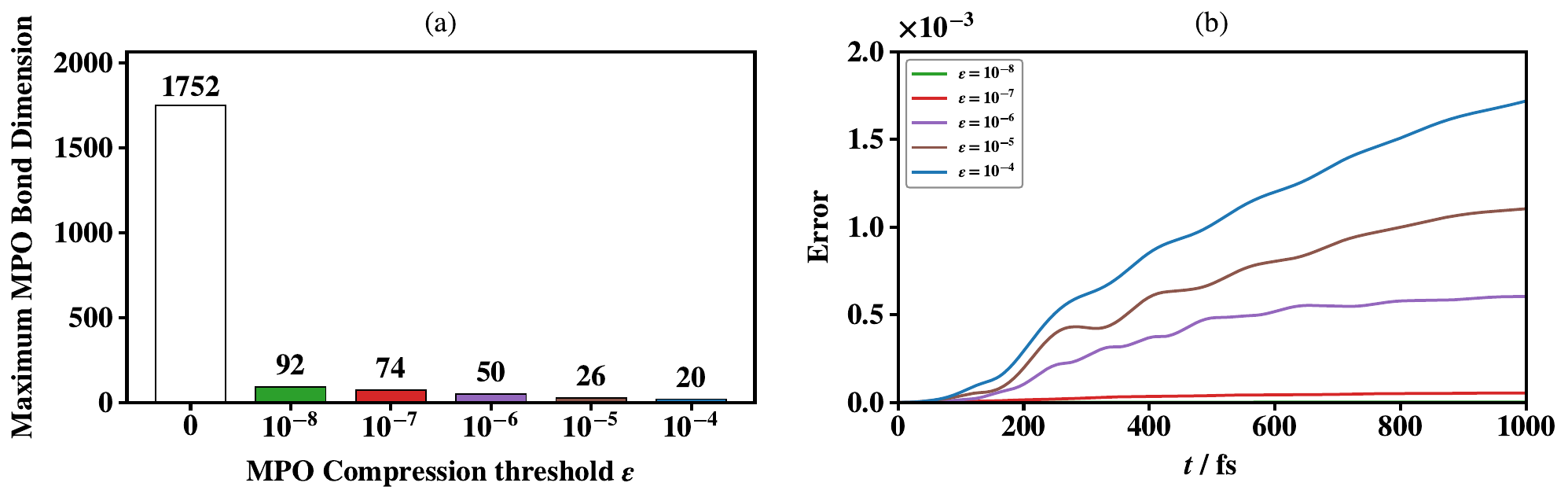}
\caption{Compressibility of the effective Hamiltonian MPO. (a) The maximum bond dimension of $\hat{H}_{\text{eff}}$ as a function of the SVD compression threshold $\varepsilon$. The bond dimension decreases drastically from the exact value (1752) to compact sizes ($<100$) with non-zero thresholds. (b) The time-dependent calculation error introduced by compression, relative to the uncompressed case. The error remains negligible (order of $10^{-3}$) even for aggressive compression ($\varepsilon=10^{-4}$).}
\label{mpo_compress}
\end{figure}

A key feature of EH-TEMPO is the compressibility of the effective Hamiltonian.
This allows the MPO representation of $\hat{H}_{\text{eff}}$ to be compressed to save computational cost without significant loss of accuracy.
Fig.~\ref{mpo_compress} demonstrates this behavior. In panel (a), without compression ($\varepsilon=0$), the bond dimension of MPO grows to 1752. By imposing an SVD truncation threshold $\varepsilon$, the bond dimension is drastically reduced (e.g., to $20$ for $\varepsilon=10^{-4}$).
Panel (b) shows that this aggressive compression incurs a minimal error penalty: even at $\varepsilon=10^{-4}$, the deviation from the uncompressed result remains negligible, with errors maintained on the order of $10^{-3}$.

\begin{figure}[htbp]
\centering
\includegraphics[width=14cm]{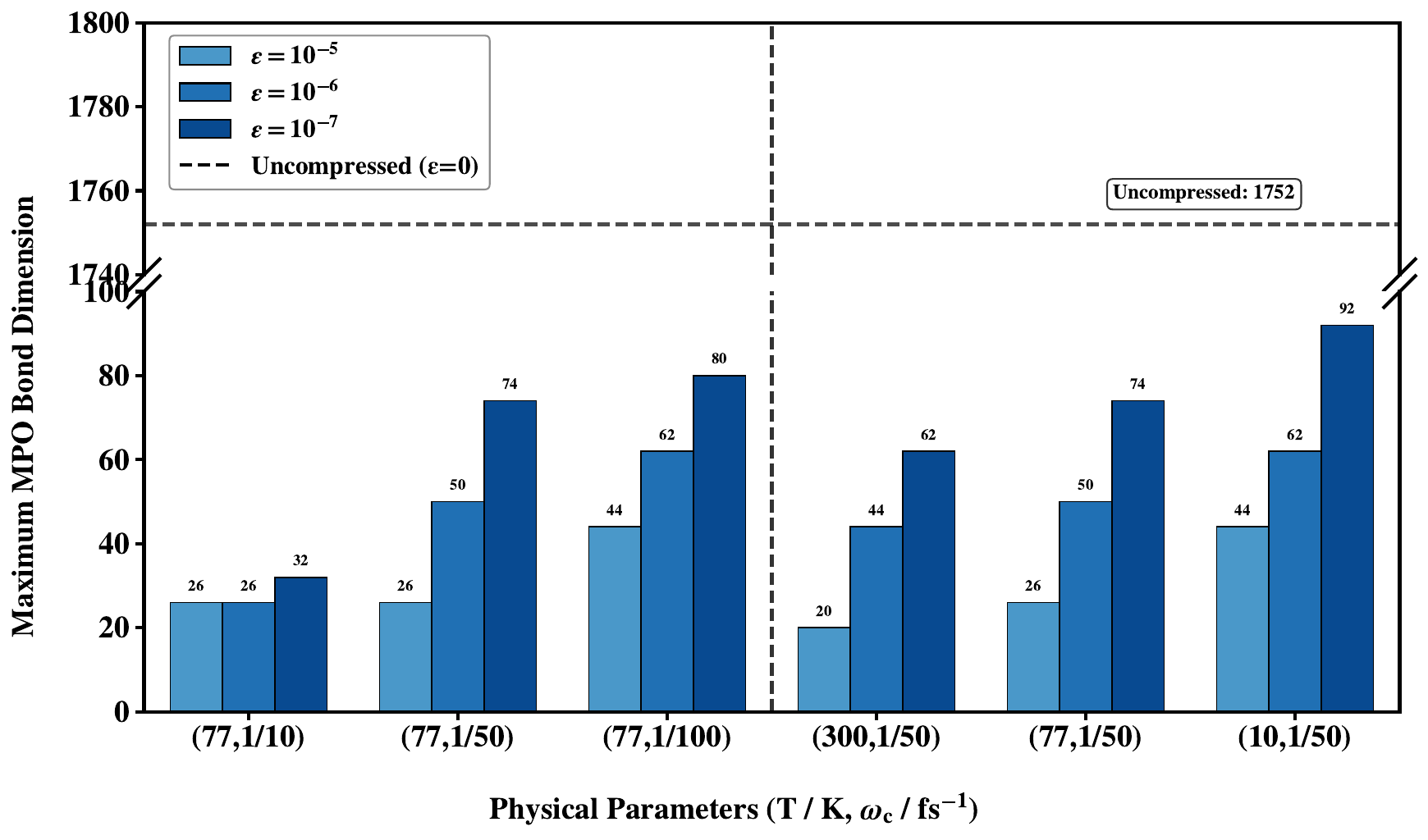}
\caption{ Comparison of the maximum bond dimension of the effective Hamiltonian MPO with various physical parameters. The x-axis labels represent the parameter pairs $(T/\text{K}, \omega_c/\text{fs}^{-1})$. The dashed line indicates the uncompressed bond dimension (1752), which is constant across these parameters for the given discretization. The colored bars show the significantly reduced bond dimensions after compression with thresholds $\varepsilon=10^{-5}$, $10^{-6}$ and $10^{-7}$, highlighting the efficiency of the compression  across different temperatures and cutoff frequencies.}
\label{max_mpo}
\end{figure}

The compressibility is further analyzed in Fig.~\ref{max_mpo} across different parameter regimes.
As expected, slower bath dynamics (left panel, $\omega_c$ from $1/10 \, \textrm{fs}^{-1}$ to $1/100 \, \textrm{fs}^{-1}$) and lower temperatures (right panel, $T$ from $300\,\text{K}$ to $\, 10 \text{K}$) lead to larger MPO bond dimensions, reflecting stronger memory effect.
Nevertheless, in all cases the compressed bond dimensions are orders of magnitude smaller than the uncompressed values (dashed lines), highlighting the effectiveness of compressing the effective Hamiltonian.
\FloatBarrier

\begin{figure}[htbp]
\centering
\includegraphics[width=0.7\textwidth]{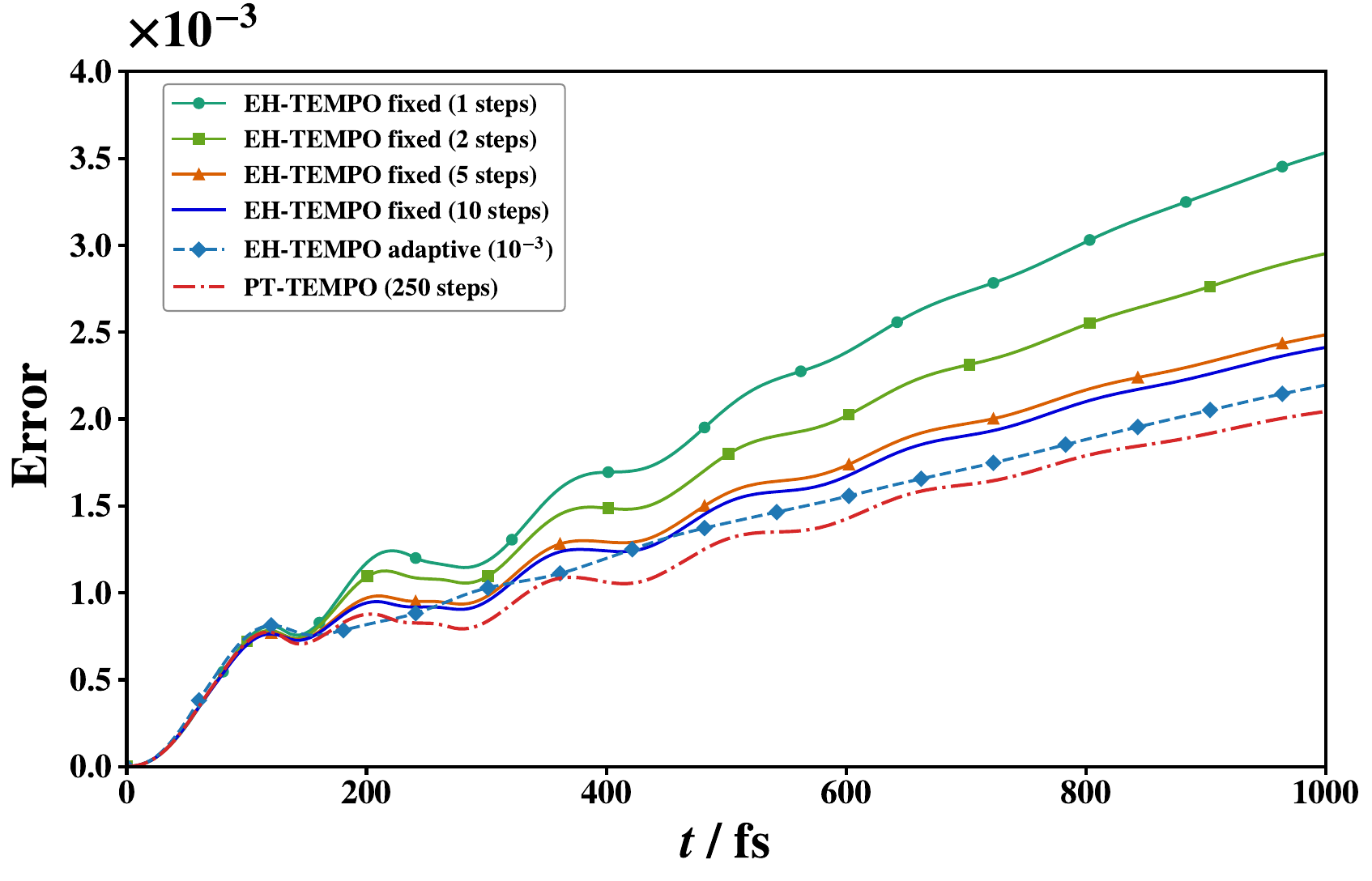}
\caption{Convergence analysis of EH-TEMPO with respect to imaginary time evolution step size. The cumulative error $\mathcal{E}(t)$ is shown for EH-TEMPO using fixed step sizes ($N_\tau = 1, 2, 5, 10$) and an adaptive step-size strategy (relative tolerance $10^{-3}$), compared against the PT-TEMPO reference ($N_t=250$). Notably, even a single evolution step ($N_\tau=1$) yields reasonable accuracy, demonstrating the efficiency of the global evolution approach. All calculations used a bond dimension of $M_S=60$.}
\label{tempo10piite}
\end{figure}

We next assess the convergence of the ITE in Fig.~\ref{tempo10piite} with different step sizes.
For fixed ITE step sizes $d\tau = 1,\, 0.5,\, 0.2,\, 0.1 / j$ ($N_\tau = 1, 2, 5, 10$, respectively), the error decreases with decreasing step size, as expected for the TDVP projector splitting algorithm.
Notably, even with $d\tau=1/j$ ($N_\tau=1$), the error remains well controlled at the same order of magnitude ($10^{-3}$).
Furthermore, the adaptive step-size algorithm can be employed: with a relative tolerance of $10^{-3}$ and an initial step size $d\tau=0.01/j$, this algorithm achieves the target accuracy more robustly than fixed-step approaches. This greatly facilitates the investigation of unexplored problems.
In all cases, EH-TEMPO requires far fewer effective evolution steps to construct the full influence functional compared to the layer-by-layer growth required by PT-TEMPO.

\begin{figure}[htbp]
\centering
\includegraphics[width=0.7\textwidth]{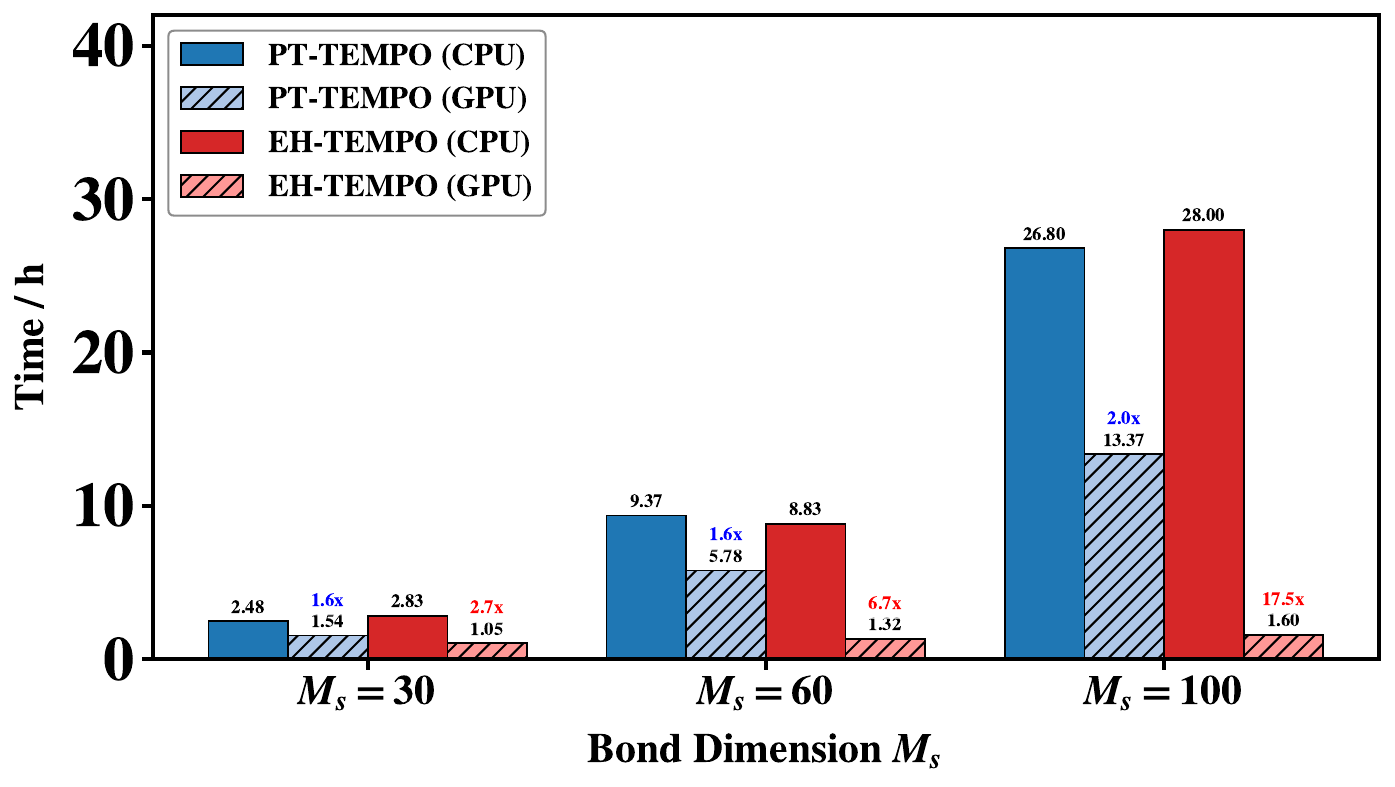}
\caption{Computational wall-clock time for the PT-TEMPO and EH-TEMPO algorithms on CPU and GPU platforms as a function of the MPS bond dimension $M_S$. The numbers above the bars indicate the specific speedup factor achieved by the GPU implementation relative to the CPU. While CPU performance is comparable between the two methods, EH-TEMPO exhibits superior scalability on GPU hardware, achieving a dramatic $17.5\times$ speedup at $M_S=100$ due to free of large tensor decomposition, compared to the modest gains observed for PT-TEMPO.}
\label{time_tempo_piite}
\end{figure}

Finally, we compare the wall-clock times of EH-TEMPO and PT-TEMPO in Fig.~\ref{time_tempo_piite}.
With 4 CPU cores (AMD EPYC 7B13), the computational cost of EH-TEMPO is comparable to that of PT-TEMPO.
It should be noted that in the PT-TEMPO implementation, the MPO of the incremental influence functional, whose bond dimension $M_O$ would be $7^2=49$, is decomposed into 7 separate layers of MPOs, each representing one independent bath with $M_O=3$ (See Appendix~\ref{appendix:scaling}).
This decomposition significantly accelerates the PT-TEMPO calculation; without it, simulations with $M_S=100$ would be computationally prohibitive.
However, since EH-TEMPO avoids large tensor decompositions, it is exceptionally well suited for GPU acceleration.
On an NVIDIA A100 GPU, EH-TEMPO achieves speedups of $2.7\times$, $6.7\times$, and $17.5\times$ for bond dimensions $M_S=30$, $60$, and $100$, respectively.
In contrast, PT-TEMPO, which relies heavily on SVD and QR decompositions---operations that are inherently difficult to parallelize on GPUs---shows limited benefit from GPU acceleration.
With GPU, EH-TEMPO can be one order of magnitude faster than PT-TEMPO for $M_S=100$.
This makes EH-TEMPO the superior choice for high-accuracy simulations requiring large bond dimensions.

\jr{
\subsection{PBI model}
\label{subsec:pbi_model}

To further demonstrate the robustness of the EH-TEMPO algorithm, particularly for benchmarking against systems with larger reorganization energies and more structured environments, we investigate the excitation energy transfer dynamics in a 4-site perylene bisimide aggregate model.

In contrast to the weakly coupled FMO complex, the PBI model places the system in a notoriously challenging intermediate-to-strong coupling regime. The environment is explicitly modeled using 28 discrete undamped intramolecular modes rather than a continuous, heavily damped spectral density, with parameters adopted from Ref.~\onlinecite{ambrosek2012quantum}. This structured, undamped bath is further complicated by a substantial reorganization energy of $\lambda = 879.1 \text{ cm}^{-1}$, relative to a nearest-neighbor excitonic coupling of $J = -500 \text{ cm}^{-1}$. This choice of environment represents a highly demanding test case; the absence of bath relaxation prevents the decay of the correlation function, leading to infinitely long-lived memory effects that typically cause rapid exponential scaling or convergence failures in standard path-integral-based methods. 

Here, we used EH-TEMPO to simulate the dynamics. The time step size is $\Delta t = 1 \text{ fs}$. The temperature is 300 K. To evaluate the accuracy of EH-TEMPO, the results are compared against numerically exact reference data generated using the TD-DMRG approach (propagating the full system-bath wavefunction with a bond dimension of $M_S=64$). Figure~\ref{fig:pbi_dynamics}(a) illustrates the convergence of the population at the initial excitation site, $n_0(t)$, with different bond dimension $M_S$. Because the memory effect is extremely strong, small bond dimensions (e.g., $M_S = 64$ to $256$) lack the capacity to capture the persistent temporal correlations, yielding trajectories that deviate significantly from the TD-DMRG reference. As $M_S$ is systematically increased, the EH-TEMPO trajectories smoothly converge. At highly expanded bond dimensions ($M_S = 1024$ and $1536$), the EH-TEMPO dynamics overlap with the numerically exact TD-DMRG reference.

This application underscores a critical advantage of the EH-TEMPO algorithm. While realistic systems with large reorganization energies and highly structured environmental modes inevitably require much larger bond dimensions to converge than the weakly coupled FMO model, reaching these necessary dimensions ($M_S > 1000$) is computationally prohibitive for standard TEMPO due to its bottleneck in tensor decomposition. EH-TEMPO, by transforming the problem into a "one-shot" imaginary time evolution and utilizing highly parallelizable tensor contractions on GPUs, makes accessing these extreme bond dimensions feasible.

Notably, this model has been successfully solved using the modular path integral approach developed by Makri and coworkers, which mitigates exponential scaling of QuAPI by partitioning the multiple baths into independent modules that are evaluated sequentially.~\cite{kundu2020modular,kundu2021efficient}  EH-TEMPO can leverage similar ideas to accelerate computations for larger aggregates: local baths can be evaluated via EH-TEMPO, and the multiple baths and system propagators can then be contracted as a 2D tensor network, akin to the strategies employed in Refs.~\onlinecite{chen2024grassmann2,bose2022multisite}.

However, it is important to note a fundamental difference in computational efficiency between EH-TEMPO and TD-DMRG in this specific undamped regime. While EH-TEMPO successfully converges, the computational cost required to reach $M_S > 1000$ is substantial. In contrast, TD-DMRG converges much faster and with a significantly smaller bond dimension ($M_S=64$). This discrepancy arises from the underlying mathematical structures of the two approaches. EH-TEMPO relies on compressing memory effects along the time axis; an undamped bath produces an infinitely long-lived correlation function, which severely penalizes the compressibility of the influence functional and demands massive bond dimensions to retain un-decaying temporal correlations. Conversely, TD-DMRG propagates the joint system-bath wavefunction explicitly. For a limited number of discrete modes over a short time, the entanglement entropy between the system and the bath modes remains relatively constrained, allowing wavefunction-based methods to capture the dynamics highly efficiently. Therefore, while EH-TEMPO successfully extends the reach of influence functional methods to the strong-coupling regime, explicit wavefunction methods like TD-DMRG are better suited for environments dominated by discrete, undamped modes.

\begin{figure}[htbp]
\centering
\subfloat[]{
    \includegraphics[width = 0.49 \textwidth]{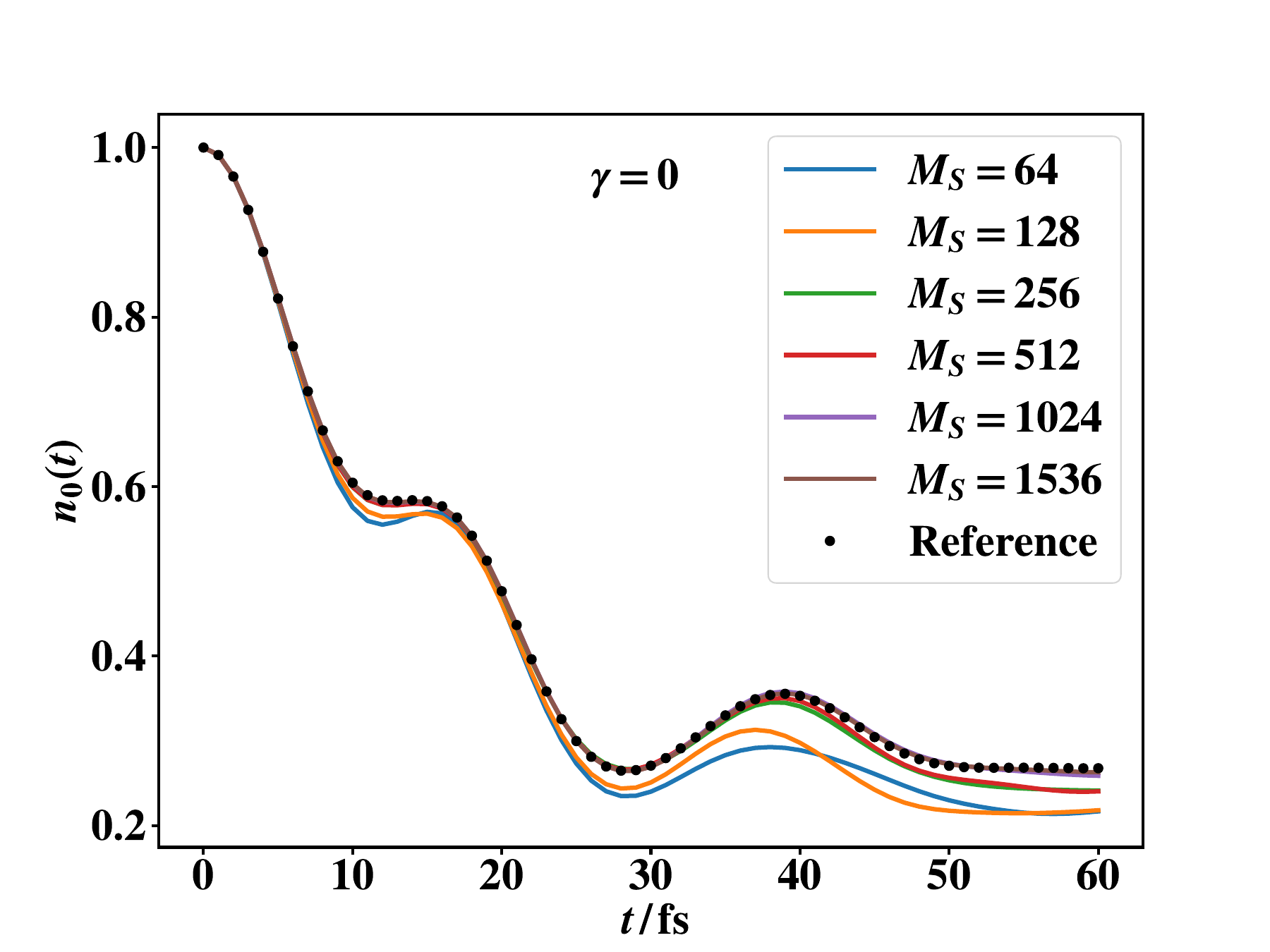}
    } 
\subfloat[]{
    \includegraphics[width = 0.49 \textwidth]{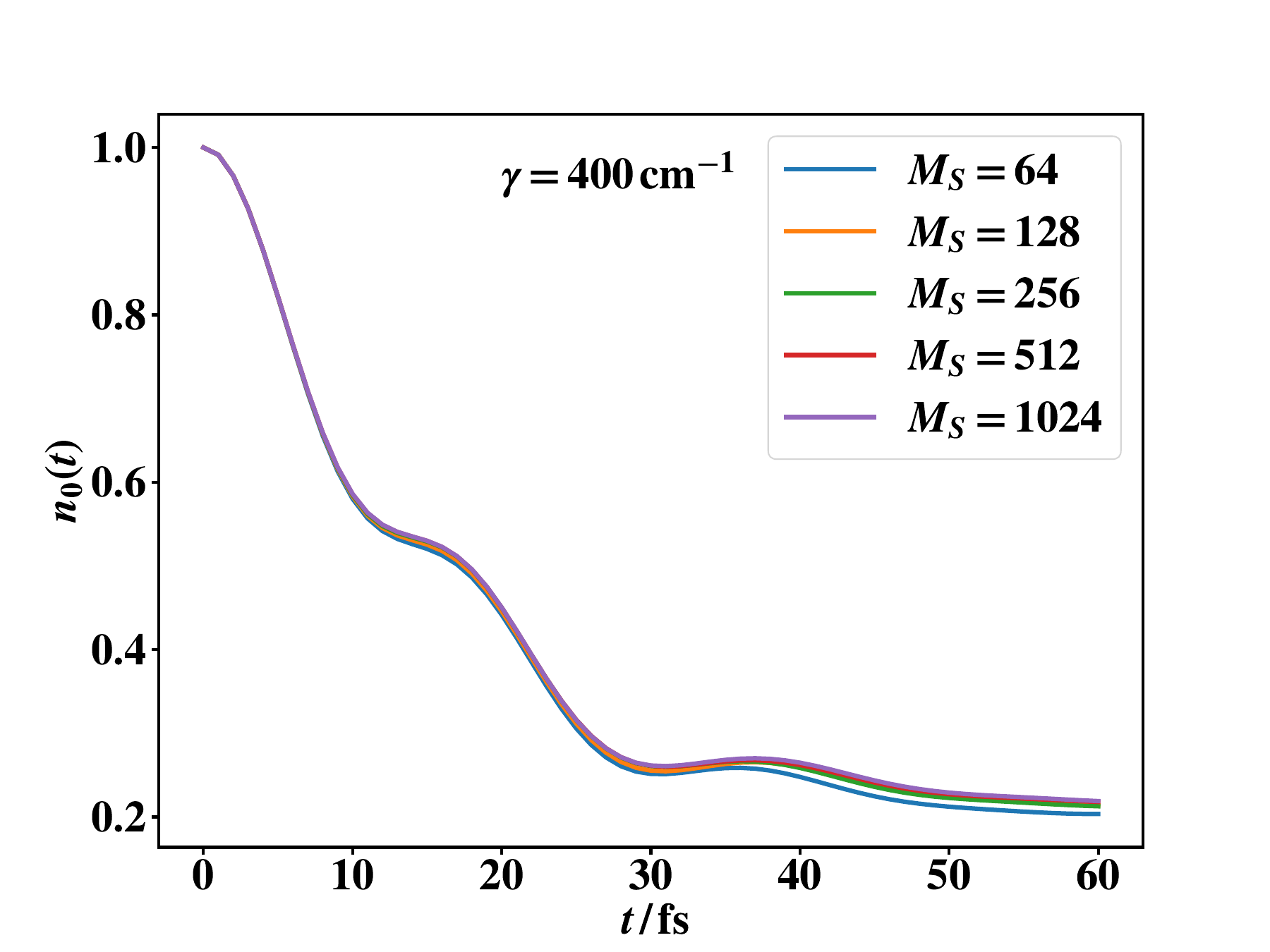}
    }   
\caption{Population dynamics $n_0(t)$ of the 4-site PBI aggregate model with (a) an undamped discrete mode bath ($\gamma = 0$) and (b) a damped Brownian bath ($\gamma = 400 \text{ cm}^{-1}$). The solid lines show the EH-TEMPO results converging as the bond dimension $M_S$ increases. The black dots represent the numerically exact TD-DMRG reference at $\gamma=0$.}
\label{fig:pbi_dynamics}
\end{figure}

In addition to the purely undamped discrete modes, we also investigated the system by replacing the environment with a Brownian oscillator spectral density, which describes the discrete modes coupled to a secondary Ohmic bath:
\begin{gather}
    J(\omega) = \sum_{i=1}^{28} \frac{2 \lambda_i \gamma_i \Omega_i ^2 \omega}{(\Omega_i^2 - \omega^2)^2 + \gamma_i^2 \omega^2},
\end{gather} 
where $\Omega_i$ is the frequency of the discrete mode with reorganization energy $\lambda_i$, and $\gamma_i$ is the damping factor. We applied a uniform damping parameter $\gamma_i = \gamma = 400 \, \text{cm}^{-1}$ across all modes. Consequently, modes with frequencies $\Omega_i < 400 \, \text{cm}^{-1}$ are overdamped, whereas those with $\Omega_i > 400 \, \text{cm}^{-1}$ are underdamped. The discrete mode case discussed above corresponds to $\gamma = 0$. The dynamics are presented in Fig.~\ref{fig:pbi_dynamics}(b). Compared to the undamped case, convergence is achieved with significantly smaller bond dimensions $M_S$. This behavior is expected, as the introduction of the damping factor accelerates the decay of the bath correlation function, thereby weakening the long-lived memory effects.
}

\section{CONCLUSION}
\label{conclusion}

In this work, we have introduced the effective Hamiltonian-based time-evolving matrix product operator (EH-TEMPO) algorithm as a highly scalable algorithm for simulating the non-Markovian dynamics of open quantum systems. The central innovation of our approach lies in reformulating the Feynman-Vernon influence functional into the imaginary time evolution of an effective Hamiltonian. This exact mathematical mapping naturally casts the effective Hamiltonian into a sum-of-products form, which seamlessly enables the use of automated, optimal MPO construction algorithms. Consequently, EH-TEMPO successfully replaces the computationally prohibitive iterative contraction-and-compression schemes of standard TEMPO and PT-TEMPO with a highly efficient projector-splitting time evolution algorithm based on time-dependent variational principle.

We benchmarked the accuracy and efficiency of EH-TEMPO on the excitation energy transfer dynamics of the 7-site Fenna-Matthews-Olson complex and a 4-site perylene bisimide aggregate. Across these diverse environments,  ranging from weakly coupled continuous Debye spectral densities to strongly coupled discrete undamped modes, EH-TEMPO accurately reproduced the numerically exact population dynamics obtained from HEOM and TD-DMRG. A pivotal finding of our analysis is the intrinsic compressibility of the effective Hamiltonian's MPO representation; we demonstrated that its bond dimensions can be aggressively truncated without compromising numerical accuracy.

A defining practical advantage of EH-TEMPO is its ability to circumvent the severe bottleneck of large tensor decompositions inherent to the original TEMPO algorithms. By relying primarily on tensor contractions, our method is exceptionally well-suited for massive GPU parallelization. In our numerical tests, EH-TEMPO achieved dramatic performance gains, delivering speedups of up to 17.5$\times$ on a single GPU for typical bond dimensions ($M_S=100$) compared to standard CPU implementations. These results indicate that EH-TEMPO is uniquely positioned to tackle complex, multi-state quantum systems that were previously computationally intractable.

\jr{
In future work, exploring the synergy between our EH-TEMPO algorithm and the SMatPI algorithm presents a highly promising avenue. Specifically, EH-TEMPO is exceptionally well-equipped to efficiently generate the influence functional data required within the memory time to initialize the SMatPI recursion. Integrating these methodologies could further alleviate memory bottlenecks, minimize tensor dimensions, and drastically reduce the overall computational scaling for simulating long-time non-Markovian dynamics in large-scale systems.
}


\acknowledgments
This work is supported by the National Natural Science
Foundation of China (Grants 22422301 and 22273005), the Quantum Science and Technology-National Science and Technology Major Project (Grant 2023ZD0300200),
and the NSAF (Grant U2330201).

\section*{Data Availability}
The data that supports the findings of this study are available within the article.

\appendix

\jr{
\section{COMPUTATIONAL COMPLEXITY ANALYSIS}
\label{appendix:scaling}

In this appendix, we provide a detailed analysis of the computational complexity, specifically tensor contraction and tensor decomposition of the (PT-)TEMPO algorithms compared to the proposed EH-TEMPO algorithm for diagonal system-bath couplings. 

For clarity, we define the following variables:
\begin{itemize}
    \item $N_t$: total number of real-time steps.
    \item $d$: system Hilbert space dimension.
    \item $M_S$: bond dimension of the MPS representing the overall influence functional.
    \item $M_O$: bond dimension of the MPO representing the incremental influence functional (in TEMPO) or the effective Hamiltonian (in EH-TEMPO).
    \item $p$: the physical dimension of the local MPS and MPO sites.
\end{itemize}
Note that the following scaling analysis considers retaining full memory without truncation.

\subsection{Standard and Optimized TEMPO Algorithms}

The standard MPO $\times$ MPS tensor contraction at a single site scales as $\mathcal{O}(M_S^2 M_O^2 p^2)$. However, if the MPO is diagonal in the physical indices, this contraction can be optimized via the Hadamard product, reducing the scaling to $\mathcal{O}(M_S^2 M_O^2 p)$. The subsequent tensor decomposition (for canonicalization and compression) of the resulting enlarged tensor scales as $\mathcal{O}(M_S^3 M_O^3 p)$.~\cite{li2020numerical}  

There are several strategies to construct the incremental influence functional in the standard TEMPO framework:

\paragraph{Grouped Path Variables (Original)} 
Following the implementation in Ref.~\onlinecite{strathearn2018efficient}, the forward and backward path variables for a single time step are grouped into a single site. This results in a physical dimension $p = d^2$. The single-layer MPO representing the incremental influence functional, $\exp\left[-\sum_{k'} (s_k^+ - s_k^-)(\eta_{kk'} s_{k'}^+ - \eta_{kk'}^* s_{k'}^-)\right]$, has a bond dimension $M_O = \mathcal{O}(d^2)$. Utilizing the diagonal property of the MPO, the contraction complexity per site is $\mathcal{O}(M_S^2 d^6)$, and the decomposition complexity is $\mathcal{O}(M_S^3 d^8)$. To simulate the dynamics over $N_t$ time steps, there are $\mathcal{O}(N_t)$ layers, and each layer contains $\mathcal{O}(N_t)$ sites. Thus, the cumulative scaling becomes $\mathcal{O}(N_t^2 M_S^2 d^6)$ for contraction and $\mathcal{O}(N_t^2 M_S^3 d^8)$ for decomposition.

\paragraph{Separated Path Variables (Optimized)} 
A significant optimization can be achieved by separating the forward and backward variables into distinct sites. This reduces the physical dimension per site to $p = d$. The incremental MPO can also be split into forward $\exp\left[-\sum_{k'} s_k^+(\eta_{kk'} s_{k'}^+ - \eta_{kk'}^* s_{k'}^-)\right]$ and backward $\exp\left[+\sum_{k'} s_k^-(\eta_{kk'} s_{k'}^+ - \eta_{kk'}^* s_{k'}^-)\right]$ components, yielding two MPOs with $M_O = \mathcal{O}(d)$. Consequently, the scaling per site is reduced to $\mathcal{O}(M_S^2 d^3)$ for contraction and $\mathcal{O}(M_S^3 d^4)$ for decomposition. Over $N_t$ time steps, the overall complexity becomes $\mathcal{O}(N_t^2 M_S^2 d^3)$ and $\mathcal{O}(N_t^2 M_S^3 d^4)$, respectively. This scaling matches the results reported in Ref.~\onlinecite{kundu2023pathsum} (Table I). Although this optimized implementation doubles both the number of sites and the number of layers, the computational scaling with respect to the system size $d$ is dramatically reduced.

The analysis above is universal for any diagonal system-bath coupling. However, for models like the Frenkel-Holstein model, the $d$ states can be represented as $d$ two-level sites, each coupled to an independent bath. Here, the incremental influence functional can be factorized into $d$ independent components. Therefore, each MPO layer can also be factorized into into $d$ sub-layers applied successively to the MPS. 
In this case, $M_O$ effectively becomes $\mathcal{O}(1)$ per sub-layer. Following the optimized TEMPO algorithm, the single-site scaling reduces to $\mathcal{O}(M_S^2 d)$ for contraction and $\mathcal{O}(M_S^3 d)$ for decomposition. Because the simulation involves $\mathcal{O}(N_t d)$ sub-layers with $\mathcal{O}(N_t)$ sites each, the total scaling over the entire simulation is $\mathcal{O}(N_t^2 M_S^2 d^2)$ for contraction and $\mathcal{O}(N_t^2 M_S^3 d^2)$ for decomposition.

\subsection{EH-TEMPO Algorithm}

EH-TEMPO fundamentally alters the computational bottleneck by framing the problem as an imaginary time evolution governed by the time-dependent variational principle, thereby avoiding the direct decomposition of heavily enlarged $M_S \times M_O$ tensors.

In the one-site TDVP-PS evolution scheme, applying the effective Hamiltonian MPO to the MPS involves projecting the action of the MPO onto the tangent space of the MPS. Generally, the contraction complexity per site is $\mathcal{O}(M_S^2 M_O^2 p^2 + M_S^3 M_O p)$.~\cite{li2020numerical} By exploiting the diagonality of the effective Hamiltonian MPO, this reduces to $\mathcal{O}(M_S^2 M_O^2 p + M_S^3 M_O p)$. Unlike standard TEMPO, updating each local site in EH-TEMPO requires $l$ tensor contractions (matrix-vector multiplications, $H \cdot c$) as demanded by the initial value problem (IVP) solver (typically $l \sim 10$). Crucially, the decomposition step within TDVP only scales as $\mathcal{O}(M_S^3 p)$ per site, because it operates directly on the MPS tensors rather than the combined MPO $\times$ MPS tensors.

Using separated path variables ($p=d$), a full TDVP sweep over the entire chain of length $\mathcal{O}(N_t)$ across a total of $N_\tau$ imaginary time evolution steps results in an overall scaling of $\mathcal{O}\big(l N_t N_\tau (M_S^2 M_O^2 d + M_S^3 M_O d)\big)$ for contraction and $\mathcal{O}(N_t N_\tau M_S^3 d)$ for decomposition.

For a single bath, the exact effective Hamiltonian MPO exhibits a bond dimension of $M_O = \mathcal{O}(N_t)$. Fortunately, as demonstrated in the main text, this effective Hamiltonian MPO is highly compressible. Regardless of whether the system is coupled to a broad continuous Debye spectral density or to discrete undamped modes, where the memory length spans the entire simulation time ($N_t \Delta t$), the actual compressed bond dimension $M_O$ becomes effectively independent of $N_t$, as illustrated in Fig.~\ref{fig:mo_fmo_pbi}. This intrinsic compressibility was similarly observed in Ref.~\onlinecite{guo2024efficient}, which employed the Prony fitting algorithm to construct the effective Hamiltonian MPO. That work revealed that $M_O$ is fundamentally governed by the number of terms required to decompose the memory coefficient $\eta_{\Delta k}$ into a sum of complex exponentials, $\eta_{\Delta k} \approx \sum_l \alpha_l \lambda_l^{\Delta k}$ (where $\alpha_l$ and $\lambda_l$ are the expansion parameters). The required number of fitting terms is significantly smaller than the total number of time steps.

\begin{figure}[htbp]
\centering
\subfloat[]{
    \includegraphics[width = 0.47 \textwidth]{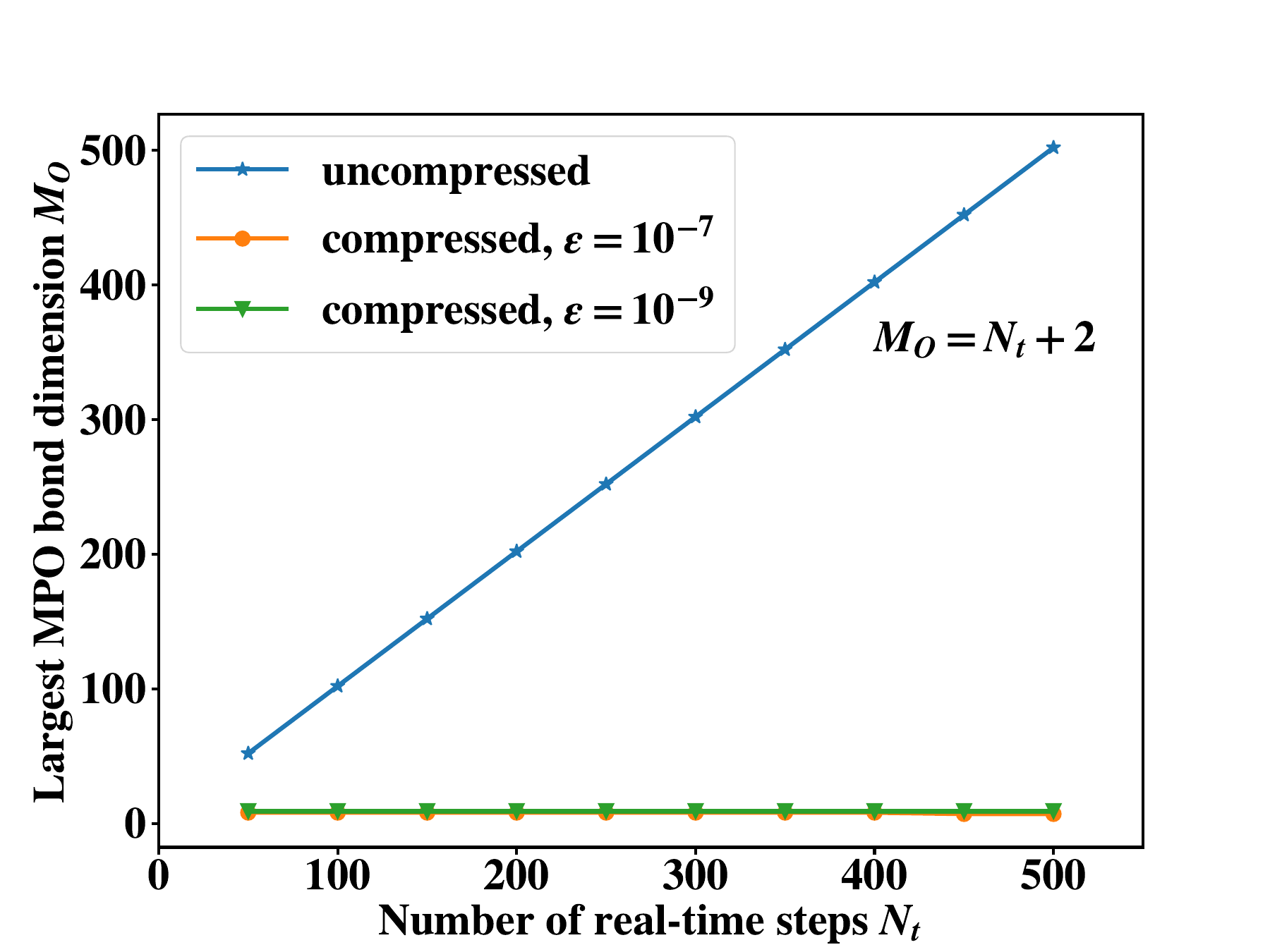}
    } 
\subfloat[]{
    \includegraphics[width = 0.47 \textwidth]{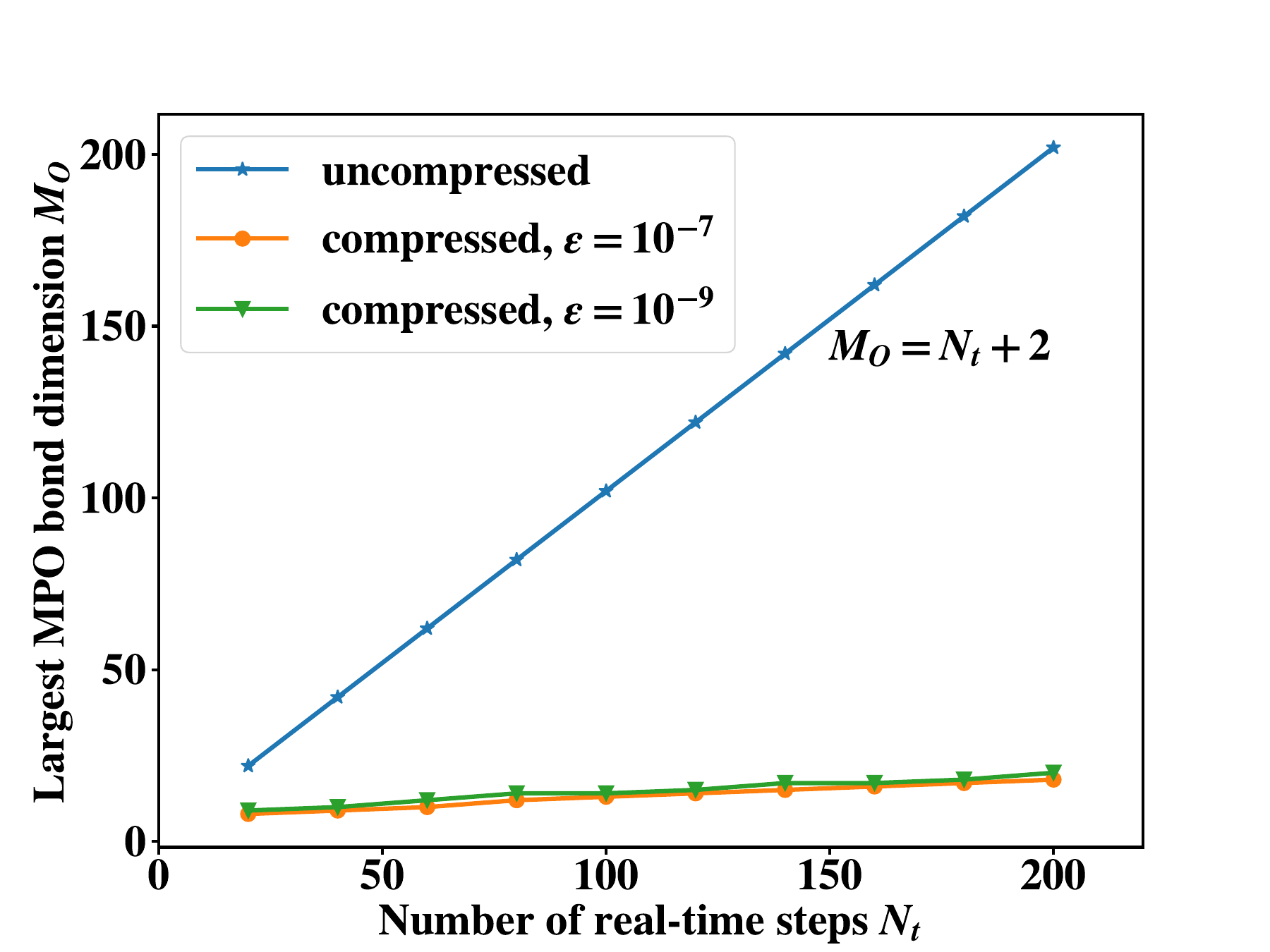}
    }
\caption{Largest uncompressed and compressed MPO bond dimensions, $M_O$, of the effective Hamiltonian for a single local bath as a function of the number of real-time steps $N_t$. Results are presented for (a) the 7-site FMO complex ($\Delta t = 4\,\text{fs}$) and (b) the 4-site PBI aggregate model ($\Delta t = 1\,\text{fs}$). Two SVD compression thresholds, $10^{-7}$ and $10^{-9}$, are compared. For the FMO model, the largest compressed bond dimensions remain constant at $M_O=8$ (for $10^{-7}$) and $M_O=9$ (for $10^{-9}$) across all values of $N_t$. For the PBI model, the largest compressed bond dimensions at $N_t=20$ are $M_O=8$ ($10^{-7}$) and $M_O=9$ ($10^{-9}$); at $N_t=200$, these values increase only moderately to $M_O=18$ and $M_O=20$, respectively.
}
\label{fig:mo_fmo_pbi}
\end{figure}

For systems coupled to $d$ independent local baths, such as the Frenkel-Holstein model, the total effective Hamiltonian is a direct summation of the individual bath contributions. Consequently, while the MPO for each individual bath component can be aggressively compressed, the overall MPO blocks cannot be interleaved or further compressed across different bath sites. As a result, the total bond dimension scales additively with the number of independent baths, yielding a compressed scaling of $M_O = \mathcal{O}(d)$.

\subsection{Summary of Computational Complexity}

Table~\ref{tab:complexity_summary} summarizes the computational complexities for maintaining full memory dynamics across $N_t$ time steps. As shown, EH-TEMPO significantly mitigates the harsh tensor decomposition bottleneck associated with standard TEMPO algorithms.

\begin{table}[htbp]
\centering
\caption{Summary of overall computational complexity for maintaining full memory across $N_t$ time steps. $N_\tau$ is the number of imaginary time steps, and $l$ is the number of IVP solver iterations per site (typically $\sim 10$).}
\label{tab:complexity_summary}
\resizebox{\textwidth}{!}{
\begin{ruledtabular} 
\begin{tabular}{lcc} 
\textbf{Algorithm} & \textbf{Contraction Scaling} & \textbf{Decomposition Scaling} \\
\colrule 
Standard TEMPO & $\mathcal{O}(N_t^2 M_S^2 d^6)$ & $\mathcal{O}(N_t^2 M_S^3 d^8)$ \\
Optimized TEMPO & $\mathcal{O}(N_t^2 M_S^2 d^3)$ & $\mathcal{O}(N_t^2 M_S^3 d^4)$ \\
Optimized TEMPO (Holstein-specific) & $\mathcal{O}(N_t^2 M_S^2 d^2)$ & $\mathcal{O}(N_t^2 M_S^3 d^2)$ \\
EH-TEMPO & $\mathcal{O}\big(l N_t N_\tau (M_S^2 M_O^2 d + M_S^3 M_O d)\big)$ & $\mathcal{O}(N_t N_\tau M_S^3 d)$ \\
\end{tabular}
\end{ruledtabular} 
}
\end{table}

}

\FloatBarrier
\bibliography{reference}

\begin{thebibliography}{58}%
\makeatletter
\providecommand \@ifxundefined [1]{%
 \@ifx{#1\undefined}
}%
\providecommand \@ifnum [1]{%
 \ifnum #1\expandafter \@firstoftwo
 \else \expandafter \@secondoftwo
 \fi
}%
\providecommand \@ifx [1]{%
 \ifx #1\expandafter \@firstoftwo
 \else \expandafter \@secondoftwo
 \fi
}%
\providecommand \natexlab [1]{#1}%
\providecommand \enquote  [1]{``#1''}%
\providecommand \bibnamefont  [1]{#1}%
\providecommand \bibfnamefont [1]{#1}%
\providecommand \citenamefont [1]{#1}%
\providecommand \href@noop [0]{\@secondoftwo}%
\providecommand \href [0]{\begingroup \@sanitize@url \@href}%
\providecommand \@href[1]{\@@startlink{#1}\@@href}%
\providecommand \@@href[1]{\endgroup#1\@@endlink}%
\providecommand \@sanitize@url [0]{\catcode `\\12\catcode `\$12\catcode `\&12\catcode `\#12\catcode `\^12\catcode `\_12\catcode `\%12\relax}%
\providecommand \@@startlink[1]{}%
\providecommand \@@endlink[0]{}%
\providecommand \url  [0]{\begingroup\@sanitize@url \@url }%
\providecommand \@url [1]{\endgroup\@href {#1}{\urlprefix }}%
\providecommand \urlprefix  [0]{URL }%
\providecommand \Eprint [0]{\href }%
\providecommand \doibase [0]{http://dx.doi.org/}%
\providecommand \selectlanguage [0]{\@gobble}%
\providecommand \bibinfo  [0]{\@secondoftwo}%
\providecommand \bibfield  [0]{\@secondoftwo}%
\providecommand \translation [1]{[#1]}%
\providecommand \BibitemOpen [0]{}%
\providecommand \bibitemStop [0]{}%
\providecommand \bibitemNoStop [0]{.\EOS\space}%
\providecommand \EOS [0]{\spacefactor3000\relax}%
\providecommand \BibitemShut  [1]{\csname bibitem#1\endcsname}%
\let\auto@bib@innerbib\@empty
\bibitem [{\citenamefont {Nitzan}(2006)}]{nitzan2006chemical}%
  \BibitemOpen
  \bibfield  {author} {\bibinfo {author} {\bibfnamefont {A.}~\bibnamefont {Nitzan}},\ }\href@noop {} {\emph {\bibinfo {title} {Chemical dynamics in condensed phases: relaxation, transfer and reactions in condensed molecular systems}}}\ (\bibinfo  {publisher} {Oxford university press},\ \bibinfo {year} {2006})\BibitemShut {NoStop}%
\bibitem [{\citenamefont {Weiss}(2012)}]{weiss2012quantum}%
  \BibitemOpen
  \bibfield  {author} {\bibinfo {author} {\bibfnamefont {U.}~\bibnamefont {Weiss}},\ }\href@noop {} {\emph {\bibinfo {title} {Quantum dissipative systems}}}\ (\bibinfo  {publisher} {World Scientific},\ \bibinfo {year} {2012})\BibitemShut {NoStop}%
\bibitem [{\citenamefont {May}\ and\ \citenamefont {K{\"u}hn}(2023)}]{may2023charge}%
  \BibitemOpen
  \bibfield  {author} {\bibinfo {author} {\bibfnamefont {V.}~\bibnamefont {May}}\ and\ \bibinfo {author} {\bibfnamefont {O.}~\bibnamefont {K{\"u}hn}},\ }\href@noop {} {\emph {\bibinfo {title} {Charge and energy transfer dynamics in molecular systems}}}\ (\bibinfo  {publisher} {John Wiley \& Sons},\ \bibinfo {year} {2023})\BibitemShut {NoStop}%
\bibitem [{\citenamefont {Yan}\ \emph {et~al.}(2021)\citenamefont {Yan}, \citenamefont {Liu}, \citenamefont {Xing},\ and\ \citenamefont {Shi}}]{biologicalsystems2}%
  \BibitemOpen
  \bibfield  {author} {\bibinfo {author} {\bibfnamefont {Y.}~\bibnamefont {Yan}}, \bibinfo {author} {\bibfnamefont {Y.}~\bibnamefont {Liu}}, \bibinfo {author} {\bibfnamefont {T.}~\bibnamefont {Xing}}, \ and\ \bibinfo {author} {\bibfnamefont {Q.}~\bibnamefont {Shi}},\ }\bibfield  {title} {\enquote {\bibinfo {title} {Theoretical study of excitation energy transfer and nonlinear spectroscopy of photosynthetic light-harvesting complexes using the nonperturbative reduced dynamics method},}\ }\href@noop {} {\bibfield  {journal} {\bibinfo  {journal} {Wiley Interdiscip Rev Comput Mol Sci}\ }\textbf {\bibinfo {volume} {11}},\ \bibinfo {pages} {e1498} (\bibinfo {year} {2021})}\BibitemShut {NoStop}%
\bibitem [{\citenamefont {Jang}\ and\ \citenamefont {Mennucci}(2018)}]{jang2018delocalized}%
  \BibitemOpen
  \bibfield  {author} {\bibinfo {author} {\bibfnamefont {S.~J.}\ \bibnamefont {Jang}}\ and\ \bibinfo {author} {\bibfnamefont {B.}~\bibnamefont {Mennucci}},\ }\bibfield  {title} {\enquote {\bibinfo {title} {Delocalized excitons in natural light-harvesting complexes},}\ }\href@noop {} {\bibfield  {journal} {\bibinfo  {journal} {Rev. Mod. Phys.}\ }\textbf {\bibinfo {volume} {90}},\ \bibinfo {pages} {035003} (\bibinfo {year} {2018})}\BibitemShut {NoStop}%
\bibitem [{\citenamefont {Xu}\ \emph {et~al.}(2026)\citenamefont {Xu}, \citenamefont {Vadimov}, \citenamefont {Stockburger},\ and\ \citenamefont {Ankerhold}}]{xu2026colloquium}%
  \BibitemOpen
  \bibfield  {author} {\bibinfo {author} {\bibfnamefont {M.}~\bibnamefont {Xu}}, \bibinfo {author} {\bibfnamefont {V.}~\bibnamefont {Vadimov}}, \bibinfo {author} {\bibfnamefont {J.}~\bibnamefont {Stockburger}}, \ and\ \bibinfo {author} {\bibfnamefont {J.}~\bibnamefont {Ankerhold}},\ }\bibfield  {title} {\enquote {\bibinfo {title} {Colloquium: Simulating non-markovian dynamics in open quantum systems},}\ }\href@noop {} {\bibfield  {journal} {\bibinfo  {journal} {Rev. Mod. Phys.}\ }\textbf {\bibinfo {volume} {98}},\ \bibinfo {pages} {021002} (\bibinfo {year} {2026})}\BibitemShut {NoStop}%
\bibitem [{\citenamefont {Ma}, \citenamefont {Luo},\ and\ \citenamefont {Yao}(2018)}]{ma2018time}%
  \BibitemOpen
  \bibfield  {author} {\bibinfo {author} {\bibfnamefont {H.}~\bibnamefont {Ma}}, \bibinfo {author} {\bibfnamefont {Z.}~\bibnamefont {Luo}}, \ and\ \bibinfo {author} {\bibfnamefont {Y.}~\bibnamefont {Yao}},\ }\bibfield  {title} {\enquote {\bibinfo {title} {The time-dependent density matrix renormalisation group method},}\ }\href@noop {} {\bibfield  {journal} {\bibinfo  {journal} {Mol. Phys.}\ }\textbf {\bibinfo {volume} {116}},\ \bibinfo {pages} {854--868} (\bibinfo {year} {2018})}\BibitemShut {NoStop}%
\bibitem [{\citenamefont {Ren}\ \emph {et~al.}(2022)\citenamefont {Ren}, \citenamefont {Li}, \citenamefont {Jiang}, \citenamefont {Wang},\ and\ \citenamefont {Shuai}}]{ren2022time}%
  \BibitemOpen
  \bibfield  {author} {\bibinfo {author} {\bibfnamefont {J.}~\bibnamefont {Ren}}, \bibinfo {author} {\bibfnamefont {W.}~\bibnamefont {Li}}, \bibinfo {author} {\bibfnamefont {T.}~\bibnamefont {Jiang}}, \bibinfo {author} {\bibfnamefont {Y.}~\bibnamefont {Wang}}, \ and\ \bibinfo {author} {\bibfnamefont {Z.}~\bibnamefont {Shuai}},\ }\bibfield  {title} {\enquote {\bibinfo {title} {Time-dependent density matrix renormalization group method for quantum dynamics in complex systems},}\ }\href@noop {} {\bibfield  {journal} {\bibinfo  {journal} {Wiley Interdiscip Rev Comput Mol Sci}\ }\textbf {\bibinfo {volume} {12}},\ \bibinfo {pages} {e1614} (\bibinfo {year} {2022})}\BibitemShut {NoStop}%
\bibitem [{\citenamefont {Schr{\"o}der}\ \emph {et~al.}(2019)\citenamefont {Schr{\"o}der}, \citenamefont {Turban}, \citenamefont {Musser}, \citenamefont {Hine},\ and\ \citenamefont {Chin}}]{schroder2019tensor}%
  \BibitemOpen
  \bibfield  {author} {\bibinfo {author} {\bibfnamefont {F.~A.}\ \bibnamefont {Schr{\"o}der}}, \bibinfo {author} {\bibfnamefont {D.~H.}\ \bibnamefont {Turban}}, \bibinfo {author} {\bibfnamefont {A.~J.}\ \bibnamefont {Musser}}, \bibinfo {author} {\bibfnamefont {N.~D.}\ \bibnamefont {Hine}}, \ and\ \bibinfo {author} {\bibfnamefont {A.~W.}\ \bibnamefont {Chin}},\ }\bibfield  {title} {\enquote {\bibinfo {title} {Tensor network simulation of multi-environmental open quantum dynamics via machine learning and entanglement renormalisation},}\ }\href@noop {} {\bibfield  {journal} {\bibinfo  {journal} {Nat. Commun.}\ }\textbf {\bibinfo {volume} {10}},\ \bibinfo {pages} {1062} (\bibinfo {year} {2019})}\BibitemShut {NoStop}%
\bibitem [{\citenamefont {Li}\ \emph {et~al.}(2024)\citenamefont {Li}, \citenamefont {Ren}, \citenamefont {Yang}, \citenamefont {Wang},\ and\ \citenamefont {Shuai}}]{li2024optimal}%
  \BibitemOpen
  \bibfield  {author} {\bibinfo {author} {\bibfnamefont {W.}~\bibnamefont {Li}}, \bibinfo {author} {\bibfnamefont {J.}~\bibnamefont {Ren}}, \bibinfo {author} {\bibfnamefont {H.}~\bibnamefont {Yang}}, \bibinfo {author} {\bibfnamefont {H.}~\bibnamefont {Wang}}, \ and\ \bibinfo {author} {\bibfnamefont {Z.}~\bibnamefont {Shuai}},\ }\bibfield  {title} {\enquote {\bibinfo {title} {Optimal tree tensor network operators for tensor network simulations: Applications to open quantum systems},}\ }\href@noop {} {\bibfield  {journal} {\bibinfo  {journal} {J. Chem. Phys.}\ }\textbf {\bibinfo {volume} {161}},\ \bibinfo {pages} {054116} (\bibinfo {year} {2024})}\BibitemShut {NoStop}%
\bibitem [{\citenamefont {Wang}\ and\ \citenamefont {Thoss}(2003)}]{multilayer1}%
  \BibitemOpen
  \bibfield  {author} {\bibinfo {author} {\bibfnamefont {H.}~\bibnamefont {Wang}}\ and\ \bibinfo {author} {\bibfnamefont {M.}~\bibnamefont {Thoss}},\ }\bibfield  {title} {\enquote {\bibinfo {title} {Multilayer formulation of the multiconfiguration time-dependent hartree theory},}\ }\href@noop {} {\bibfield  {journal} {\bibinfo  {journal} {J. Chem. Phys.}\ }\textbf {\bibinfo {volume} {119}},\ \bibinfo {pages} {1289--1299} (\bibinfo {year} {2003})}\BibitemShut {NoStop}%
\bibitem [{\citenamefont {Wang}(2015)}]{wang2015multilayer}%
  \BibitemOpen
  \bibfield  {author} {\bibinfo {author} {\bibfnamefont {H.}~\bibnamefont {Wang}},\ }\bibfield  {title} {\enquote {\bibinfo {title} {Multilayer multiconfiguration time-dependent hartree theory},}\ }\href@noop {} {\bibfield  {journal} {\bibinfo  {journal} {J. Phys. Chem. A}\ }\textbf {\bibinfo {volume} {119}},\ \bibinfo {pages} {7951--7965} (\bibinfo {year} {2015})}\BibitemShut {NoStop}%
\bibitem [{\citenamefont {Lindblad}(1976)}]{lindblad1}%
  \BibitemOpen
  \bibfield  {author} {\bibinfo {author} {\bibfnamefont {G.}~\bibnamefont {Lindblad}},\ }\bibfield  {title} {\enquote {\bibinfo {title} {On the generators of quantum dynamical semigroups},}\ }\href@noop {} {\bibfield  {journal} {\bibinfo  {journal} {Communications in mathematical physics}\ }\textbf {\bibinfo {volume} {48}},\ \bibinfo {pages} {119--130} (\bibinfo {year} {1976})}\BibitemShut {NoStop}%
\bibitem [{\citenamefont {Gorini}, \citenamefont {Kossakowski},\ and\ \citenamefont {Sudarshan}(1976)}]{gorini1976completely}%
  \BibitemOpen
  \bibfield  {author} {\bibinfo {author} {\bibfnamefont {V.}~\bibnamefont {Gorini}}, \bibinfo {author} {\bibfnamefont {A.}~\bibnamefont {Kossakowski}}, \ and\ \bibinfo {author} {\bibfnamefont {E.~C.~G.}\ \bibnamefont {Sudarshan}},\ }\bibfield  {title} {\enquote {\bibinfo {title} {Completely positive dynamical semigroups of n-level systems},}\ }\href@noop {} {\bibfield  {journal} {\bibinfo  {journal} {Journal of Mathematical Physics}\ }\textbf {\bibinfo {volume} {17}},\ \bibinfo {pages} {821--825} (\bibinfo {year} {1976})}\BibitemShut {NoStop}%
\bibitem [{\citenamefont {Redfield}(1957)}]{redfield1957theory}%
  \BibitemOpen
  \bibfield  {author} {\bibinfo {author} {\bibfnamefont {A.~G.}\ \bibnamefont {Redfield}},\ }\bibfield  {title} {\enquote {\bibinfo {title} {On the theory of relaxation processes},}\ }\href@noop {} {\bibfield  {journal} {\bibinfo  {journal} {IBM Journal of Research and Development}\ }\textbf {\bibinfo {volume} {1}},\ \bibinfo {pages} {19--31} (\bibinfo {year} {1957})}\BibitemShut {NoStop}%
\bibitem [{\citenamefont {Redfield}(1965)}]{redfield1965theory}%
  \BibitemOpen
  \bibfield  {author} {\bibinfo {author} {\bibfnamefont {A.~G.}\ \bibnamefont {Redfield}},\ }\bibfield  {title} {\enquote {\bibinfo {title} {The theory of relaxation processes},}\ }in\ \href@noop {} {\emph {\bibinfo {booktitle} {Advances in Magnetic and Optical Resonance}}},\ Vol.~\bibinfo {volume} {1}\ (\bibinfo  {publisher} {Elsevier},\ \bibinfo {year} {1965})\ pp.\ \bibinfo {pages} {1--32}\BibitemShut {NoStop}%
\bibitem [{\citenamefont {Tanimura}\ and\ \citenamefont {Kubo}(1989)}]{HEOM1}%
  \BibitemOpen
  \bibfield  {author} {\bibinfo {author} {\bibfnamefont {Y.}~\bibnamefont {Tanimura}}\ and\ \bibinfo {author} {\bibfnamefont {R.}~\bibnamefont {Kubo}},\ }\bibfield  {title} {\enquote {\bibinfo {title} {Time evolution of a quantum system in contact with a nearly gaussian-markoffian noise bath},}\ }\href@noop {} {\bibfield  {journal} {\bibinfo  {journal} {Journal of the Physical Society of Japan}\ }\textbf {\bibinfo {volume} {58}},\ \bibinfo {pages} {101--114} (\bibinfo {year} {1989})}\BibitemShut {NoStop}%
\bibitem [{\citenamefont {Tanimura}(2020)}]{tanimura2020numerically}%
  \BibitemOpen
  \bibfield  {author} {\bibinfo {author} {\bibfnamefont {Y.}~\bibnamefont {Tanimura}},\ }\bibfield  {title} {\enquote {\bibinfo {title} {Numerically “exact” approach to open quantum dynamics: The hierarchical equations of motion (heom)},}\ }\href@noop {} {\bibfield  {journal} {\bibinfo  {journal} {J. Chem. Phys.}\ }\textbf {\bibinfo {volume} {153}},\ \bibinfo {pages} {020901} (\bibinfo {year} {2020})}\BibitemShut {NoStop}%
\bibitem [{\citenamefont {Di{\'o}si}, \citenamefont {Gisin},\ and\ \citenamefont {Strunz}(1998)}]{diosi1998non}%
  \BibitemOpen
  \bibfield  {author} {\bibinfo {author} {\bibfnamefont {L.}~\bibnamefont {Di{\'o}si}}, \bibinfo {author} {\bibfnamefont {N.}~\bibnamefont {Gisin}}, \ and\ \bibinfo {author} {\bibfnamefont {W.~T.}\ \bibnamefont {Strunz}},\ }\bibfield  {title} {\enquote {\bibinfo {title} {Non-markovian quantum state diffusion},}\ }\href@noop {} {\bibfield  {journal} {\bibinfo  {journal} {Phys. Rev. A}\ }\textbf {\bibinfo {volume} {58}},\ \bibinfo {pages} {1699} (\bibinfo {year} {1998})}\BibitemShut {NoStop}%
\bibitem [{\citenamefont {Suess}, \citenamefont {Eisfeld},\ and\ \citenamefont {Strunz}(2014)}]{suess2014hierarchy}%
  \BibitemOpen
  \bibfield  {author} {\bibinfo {author} {\bibfnamefont {D.}~\bibnamefont {Suess}}, \bibinfo {author} {\bibfnamefont {A.}~\bibnamefont {Eisfeld}}, \ and\ \bibinfo {author} {\bibfnamefont {W.~T.}\ \bibnamefont {Strunz}},\ }\bibfield  {title} {\enquote {\bibinfo {title} {Hierarchy of stochastic pure states for open quantum system dynamics},}\ }\href@noop {} {\bibfield  {journal} {\bibinfo  {journal} {Phys. Rev. Lett.}\ }\textbf {\bibinfo {volume} {113}},\ \bibinfo {pages} {150403} (\bibinfo {year} {2014})}\BibitemShut {NoStop}%
\bibitem [{\citenamefont {Makri}\ and\ \citenamefont {Makarov}(1995{\natexlab{a}})}]{makri1995tensor}%
  \BibitemOpen
  \bibfield  {author} {\bibinfo {author} {\bibfnamefont {N.}~\bibnamefont {Makri}}\ and\ \bibinfo {author} {\bibfnamefont {D.~E.}\ \bibnamefont {Makarov}},\ }\bibfield  {title} {\enquote {\bibinfo {title} {Tensor propagator for iterative quantum time evolution of reduced density matrices. i. theory},}\ }\href@noop {} {\bibfield  {journal} {\bibinfo  {journal} {J. Chem. Phys.}\ }\textbf {\bibinfo {volume} {102}},\ \bibinfo {pages} {4600--4610} (\bibinfo {year} {1995}{\natexlab{a}})}\BibitemShut {NoStop}%
\bibitem [{\citenamefont {Makri}\ and\ \citenamefont {Makarov}(1995{\natexlab{b}})}]{makri1995tensor2}%
  \BibitemOpen
  \bibfield  {author} {\bibinfo {author} {\bibfnamefont {N.}~\bibnamefont {Makri}}\ and\ \bibinfo {author} {\bibfnamefont {D.~I.}\ \bibnamefont {Makarov}},\ }\bibfield  {title} {\enquote {\bibinfo {title} {Tensor propagator for iterative quantum time evolution of reduced density matrices. ii. numerical methodology},}\ }\href@noop {} {\bibfield  {journal} {\bibinfo  {journal} {J. Chem. Phys.}\ }\textbf {\bibinfo {volume} {102}},\ \bibinfo {pages} {4611--4618} (\bibinfo {year} {1995}{\natexlab{b}})}\BibitemShut {NoStop}%
\bibitem [{\citenamefont {Makri}(1992)}]{makri1992improved}%
  \BibitemOpen
  \bibfield  {author} {\bibinfo {author} {\bibfnamefont {N.}~\bibnamefont {Makri}},\ }\bibfield  {title} {\enquote {\bibinfo {title} {Improved feynman propagators on a grid and non-adiabatic corrections within the path integral framework},}\ }\href@noop {} {\bibfield  {journal} {\bibinfo  {journal} {Chem. Phys. Lett.}\ }\textbf {\bibinfo {volume} {193}},\ \bibinfo {pages} {435--445} (\bibinfo {year} {1992})}\BibitemShut {NoStop}%
\bibitem [{\citenamefont {Makri}(1998)}]{makri1998quantum}%
  \BibitemOpen
  \bibfield  {author} {\bibinfo {author} {\bibfnamefont {N.}~\bibnamefont {Makri}},\ }\bibfield  {title} {\enquote {\bibinfo {title} {Quantum dissipative dynamics: A numerically exact methodology},}\ }\href@noop {} {\bibfield  {journal} {\bibinfo  {journal} {J. Phys. Chem. A}\ }\textbf {\bibinfo {volume} {102}},\ \bibinfo {pages} {4414--4427} (\bibinfo {year} {1998})}\BibitemShut {NoStop}%
\bibitem [{\citenamefont {Sim}(2001)}]{sim2001quantum}%
  \BibitemOpen
  \bibfield  {author} {\bibinfo {author} {\bibfnamefont {E.}~\bibnamefont {Sim}},\ }\bibfield  {title} {\enquote {\bibinfo {title} {Quantum dynamics for a system coupled to slow baths: On-the-fly filtered propagator method},}\ }\href@noop {} {\bibfield  {journal} {\bibinfo  {journal} {J. Chem. Phys.}\ }\textbf {\bibinfo {volume} {115}},\ \bibinfo {pages} {4450--4456} (\bibinfo {year} {2001})}\BibitemShut {NoStop}%
\bibitem [{\citenamefont {Makri}(2017)}]{makri2017iterative}%
  \BibitemOpen
  \bibfield  {author} {\bibinfo {author} {\bibfnamefont {N.}~\bibnamefont {Makri}},\ }\bibfield  {title} {\enquote {\bibinfo {title} {Iterative blip-summed path integral for quantum dynamics in strongly dissipative environments},}\ }\href@noop {} {\bibfield  {journal} {\bibinfo  {journal} {J. Chem. Phys.}\ }\textbf {\bibinfo {volume} {146}},\ \bibinfo {pages} {134101} (\bibinfo {year} {2017})}\BibitemShut {NoStop}%
\bibitem [{\citenamefont {Ovcharenko}, \citenamefont {Xu},\ and\ \citenamefont {Fingerhut}(2026)}]{Ovcharenko2026Scalable}%
  \BibitemOpen
  \bibfield  {author} {\bibinfo {author} {\bibfnamefont {R.}~\bibnamefont {Ovcharenko}}, \bibinfo {author} {\bibfnamefont {X.}~\bibnamefont {Xu}}, \ and\ \bibinfo {author} {\bibfnamefont {B.~P.}\ \bibnamefont {Fingerhut}},\ }\bibfield  {title} {\enquote {\bibinfo {title} {Scalable distributed memory implementation of the quasi-adiabatic propagator path integral},}\ }\href@noop {} {\bibfield  {journal} {\bibinfo  {journal} {J. Chem. Theory Comput.}\ }\textbf {\bibinfo {volume} {22}},\ \bibinfo {pages} {3810--3826} (\bibinfo {year} {2026})}\BibitemShut {NoStop}%
\bibitem [{\citenamefont {Makri}(2020)}]{makri2020small}%
  \BibitemOpen
  \bibfield  {author} {\bibinfo {author} {\bibfnamefont {N.}~\bibnamefont {Makri}},\ }\bibfield  {title} {\enquote {\bibinfo {title} {Small matrix disentanglement of the path integral: overcoming the exponential tensor scaling with memory length},}\ }\href@noop {} {\bibfield  {journal} {\bibinfo  {journal} {J. Chem. Phys.}\ }\textbf {\bibinfo {volume} {152}} (\bibinfo {year} {2020})}\BibitemShut {NoStop}%
\bibitem [{\citenamefont {Strathearn}\ \emph {et~al.}(2018)\citenamefont {Strathearn}, \citenamefont {Kirton}, \citenamefont {Kilda}, \citenamefont {Keeling},\ and\ \citenamefont {Lovett}}]{strathearn2018efficient}%
  \BibitemOpen
  \bibfield  {author} {\bibinfo {author} {\bibfnamefont {A.}~\bibnamefont {Strathearn}}, \bibinfo {author} {\bibfnamefont {P.}~\bibnamefont {Kirton}}, \bibinfo {author} {\bibfnamefont {D.}~\bibnamefont {Kilda}}, \bibinfo {author} {\bibfnamefont {J.}~\bibnamefont {Keeling}}, \ and\ \bibinfo {author} {\bibfnamefont {B.~W.}\ \bibnamefont {Lovett}},\ }\bibfield  {title} {\enquote {\bibinfo {title} {Efficient non-markovian quantum dynamics using time-evolving matrix product operators},}\ }\href@noop {} {\bibfield  {journal} {\bibinfo  {journal} {Nat. Commun.}\ }\textbf {\bibinfo {volume} {9}},\ \bibinfo {pages} {3322} (\bibinfo {year} {2018})}\BibitemShut {NoStop}%
\bibitem [{\citenamefont {Ye}\ and\ \citenamefont {Chan}(2021)}]{ye2021constructing}%
  \BibitemOpen
  \bibfield  {author} {\bibinfo {author} {\bibfnamefont {E.}~\bibnamefont {Ye}}\ and\ \bibinfo {author} {\bibfnamefont {G.~K.-L.}\ \bibnamefont {Chan}},\ }\bibfield  {title} {\enquote {\bibinfo {title} {Constructing tensor network influence functionals for general quantum dynamics},}\ }\href@noop {} {\bibfield  {journal} {\bibinfo  {journal} {J. Chem. Phys.}\ }\textbf {\bibinfo {volume} {155}},\ \bibinfo {pages} {044104} (\bibinfo {year} {2021})}\BibitemShut {NoStop}%
\bibitem [{\citenamefont {Cygorek}\ \emph {et~al.}(2022)\citenamefont {Cygorek}, \citenamefont {Cosacchi}, \citenamefont {Vagov}, \citenamefont {Axt}, \citenamefont {Lovett}, \citenamefont {Keeling},\ and\ \citenamefont {Gauger}}]{cygorek2022simulation}%
  \BibitemOpen
  \bibfield  {author} {\bibinfo {author} {\bibfnamefont {M.}~\bibnamefont {Cygorek}}, \bibinfo {author} {\bibfnamefont {M.}~\bibnamefont {Cosacchi}}, \bibinfo {author} {\bibfnamefont {A.}~\bibnamefont {Vagov}}, \bibinfo {author} {\bibfnamefont {V.~M.}\ \bibnamefont {Axt}}, \bibinfo {author} {\bibfnamefont {B.~W.}\ \bibnamefont {Lovett}}, \bibinfo {author} {\bibfnamefont {J.}~\bibnamefont {Keeling}}, \ and\ \bibinfo {author} {\bibfnamefont {E.~M.}\ \bibnamefont {Gauger}},\ }\bibfield  {title} {\enquote {\bibinfo {title} {Simulation of open quantum systems by automated compression of arbitrary environments},}\ }\href@noop {} {\bibfield  {journal} {\bibinfo  {journal} {Nat. Phys.}\ }\textbf {\bibinfo {volume} {18}},\ \bibinfo {pages} {662--668} (\bibinfo {year} {2022})}\BibitemShut {NoStop}%
\bibitem [{\citenamefont {Schollw{\"o}ck}(2011)}]{schollwock2011density}%
  \BibitemOpen
  \bibfield  {author} {\bibinfo {author} {\bibfnamefont {U.}~\bibnamefont {Schollw{\"o}ck}},\ }\bibfield  {title} {\enquote {\bibinfo {title} {The density-matrix renormalization group in the age of matrix product states},}\ }\href@noop {} {\bibfield  {journal} {\bibinfo  {journal} {Ann. Phys.}\ }\textbf {\bibinfo {volume} {326}},\ \bibinfo {pages} {96--192} (\bibinfo {year} {2011})}\BibitemShut {NoStop}%
\bibitem [{\citenamefont {J{\o}rgensen}\ and\ \citenamefont {Pollock}(2019)}]{jorgensen2019exploiting}%
  \BibitemOpen
  \bibfield  {author} {\bibinfo {author} {\bibfnamefont {M.~R.}\ \bibnamefont {J{\o}rgensen}}\ and\ \bibinfo {author} {\bibfnamefont {F.~A.}\ \bibnamefont {Pollock}},\ }\bibfield  {title} {\enquote {\bibinfo {title} {Exploiting the causal tensor network structure of quantum processes to efficiently simulate non-markovian path integrals},}\ }\href@noop {} {\bibfield  {journal} {\bibinfo  {journal} {Phys. Rev. Lett.}\ }\textbf {\bibinfo {volume} {123}},\ \bibinfo {pages} {240602} (\bibinfo {year} {2019})}\BibitemShut {NoStop}%
\bibitem [{\citenamefont {Richter}\ and\ \citenamefont {Hughes}(2022)}]{richter2022enhanced}%
  \BibitemOpen
  \bibfield  {author} {\bibinfo {author} {\bibfnamefont {M.}~\bibnamefont {Richter}}\ and\ \bibinfo {author} {\bibfnamefont {S.}~\bibnamefont {Hughes}},\ }\bibfield  {title} {\enquote {\bibinfo {title} {Enhanced tempo algorithm for quantum path integrals with off-diagonal system-bath coupling: Applications to photonic quantum networks},}\ }\href@noop {} {\bibfield  {journal} {\bibinfo  {journal} {Phys. Rev. Lett.}\ }\textbf {\bibinfo {volume} {128}},\ \bibinfo {pages} {167403} (\bibinfo {year} {2022})}\BibitemShut {NoStop}%
\bibitem [{\citenamefont {Gribben}\ \emph {et~al.}(2022)\citenamefont {Gribben}, \citenamefont {Rouse}, \citenamefont {Iles-Smith}, \citenamefont {Strathearn}, \citenamefont {Maguire}, \citenamefont {Kirton}, \citenamefont {Nazir}, \citenamefont {Gauger},\ and\ \citenamefont {Lovett}}]{gribben2022exact}%
  \BibitemOpen
  \bibfield  {author} {\bibinfo {author} {\bibfnamefont {D.}~\bibnamefont {Gribben}}, \bibinfo {author} {\bibfnamefont {D.~M.}\ \bibnamefont {Rouse}}, \bibinfo {author} {\bibfnamefont {J.}~\bibnamefont {Iles-Smith}}, \bibinfo {author} {\bibfnamefont {A.}~\bibnamefont {Strathearn}}, \bibinfo {author} {\bibfnamefont {H.}~\bibnamefont {Maguire}}, \bibinfo {author} {\bibfnamefont {P.}~\bibnamefont {Kirton}}, \bibinfo {author} {\bibfnamefont {A.}~\bibnamefont {Nazir}}, \bibinfo {author} {\bibfnamefont {E.~M.}\ \bibnamefont {Gauger}}, \ and\ \bibinfo {author} {\bibfnamefont {B.~W.}\ \bibnamefont {Lovett}},\ }\bibfield  {title} {\enquote {\bibinfo {title} {Exact dynamics of nonadditive environments in non-markovian open quantum systems},}\ }\href@noop {} {\bibfield  {journal} {\bibinfo  {journal} {PRX Quantum}\ }\textbf {\bibinfo {volume} {3}},\ \bibinfo {pages} {010321} (\bibinfo {year} {2022})}\BibitemShut {NoStop}%
\bibitem [{\citenamefont {Chen}(2024)}]{chen2024path}%
  \BibitemOpen
  \bibfield  {author} {\bibinfo {author} {\bibfnamefont {R.}~\bibnamefont {Chen}},\ }\bibfield  {title} {\enquote {\bibinfo {title} {Path integral formalism of open quantum systems with non-diagonal system-bath coupling},}\ }\href@noop {} {\bibfield  {journal} {\bibinfo  {journal} {Communications in Theoretical Physics}\ }\textbf {\bibinfo {volume} {76}},\ \bibinfo {pages} {115701} (\bibinfo {year} {2024})}\BibitemShut {NoStop}%
\bibitem [{\citenamefont {Zhang}\ and\ \citenamefont {Shi}(2025)}]{zhang2025time}%
  \BibitemOpen
  \bibfield  {author} {\bibinfo {author} {\bibfnamefont {S.}~\bibnamefont {Zhang}}\ and\ \bibinfo {author} {\bibfnamefont {Q.}~\bibnamefont {Shi}},\ }\bibfield  {title} {\enquote {\bibinfo {title} {Time-evolving matrix product operator method in a nondiagonal basis set based on the derivative of the path-integral expression},}\ }\href@noop {} {\bibfield  {journal} {\bibinfo  {journal} {Phys. Rev. A}\ }\textbf {\bibinfo {volume} {111}},\ \bibinfo {pages} {062210} (\bibinfo {year} {2025})}\BibitemShut {NoStop}%
\bibitem [{\citenamefont {Bose}(2023)}]{bose2023quantum}%
  \BibitemOpen
  \bibfield  {author} {\bibinfo {author} {\bibfnamefont {A.}~\bibnamefont {Bose}},\ }\bibfield  {title} {\enquote {\bibinfo {title} {Quantum correlation functions through tensor network path integral},}\ }\href@noop {} {\bibfield  {journal} {\bibinfo  {journal} {J. Chem. Phys.}\ }\textbf {\bibinfo {volume} {159}},\ \bibinfo {pages} {214110} (\bibinfo {year} {2023})}\BibitemShut {NoStop}%
\bibitem [{\citenamefont {Ng}\ \emph {et~al.}(2023)\citenamefont {Ng}, \citenamefont {Park}, \citenamefont {Millis}, \citenamefont {Chan},\ and\ \citenamefont {Reichman}}]{ng2023real}%
  \BibitemOpen
  \bibfield  {author} {\bibinfo {author} {\bibfnamefont {N.}~\bibnamefont {Ng}}, \bibinfo {author} {\bibfnamefont {G.}~\bibnamefont {Park}}, \bibinfo {author} {\bibfnamefont {A.~J.}\ \bibnamefont {Millis}}, \bibinfo {author} {\bibfnamefont {G.~K.-L.}\ \bibnamefont {Chan}}, \ and\ \bibinfo {author} {\bibfnamefont {D.~R.}\ \bibnamefont {Reichman}},\ }\bibfield  {title} {\enquote {\bibinfo {title} {Real-time evolution of anderson impurity models via tensor network influence functionals},}\ }\href@noop {} {\bibfield  {journal} {\bibinfo  {journal} {Phys. Rev. B}\ }\textbf {\bibinfo {volume} {107}},\ \bibinfo {pages} {125103} (\bibinfo {year} {2023})}\BibitemShut {NoStop}%
\bibitem [{\citenamefont {Chen}, \citenamefont {Xu},\ and\ \citenamefont {Guo}(2024{\natexlab{a}})}]{chen2024grassmann}%
  \BibitemOpen
  \bibfield  {author} {\bibinfo {author} {\bibfnamefont {R.}~\bibnamefont {Chen}}, \bibinfo {author} {\bibfnamefont {X.}~\bibnamefont {Xu}}, \ and\ \bibinfo {author} {\bibfnamefont {C.}~\bibnamefont {Guo}},\ }\bibfield  {title} {\enquote {\bibinfo {title} {Grassmann time-evolving matrix product operators for quantum impurity models},}\ }\href@noop {} {\bibfield  {journal} {\bibinfo  {journal} {Phys. Rev. B}\ }\textbf {\bibinfo {volume} {109}},\ \bibinfo {pages} {045140} (\bibinfo {year} {2024}{\natexlab{a}})}\BibitemShut {NoStop}%
\bibitem [{\citenamefont {Chen}, \citenamefont {Xu},\ and\ \citenamefont {Guo}(2024{\natexlab{b}})}]{chen2024grassmann2}%
  \BibitemOpen
  \bibfield  {author} {\bibinfo {author} {\bibfnamefont {R.}~\bibnamefont {Chen}}, \bibinfo {author} {\bibfnamefont {X.}~\bibnamefont {Xu}}, \ and\ \bibinfo {author} {\bibfnamefont {C.}~\bibnamefont {Guo}},\ }\bibfield  {title} {\enquote {\bibinfo {title} {Grassmann time-evolving matrix product operators for equilibrium quantum impurity problems},}\ }\href@noop {} {\bibfield  {journal} {\bibinfo  {journal} {New J. Phys.}\ }\textbf {\bibinfo {volume} {26}},\ \bibinfo {pages} {013019} (\bibinfo {year} {2024}{\natexlab{b}})}\BibitemShut {NoStop}%
\bibitem [{\citenamefont {Kundu}\ and\ \citenamefont {Makri}(2023)}]{kundu2023pathsum}%
  \BibitemOpen
  \bibfield  {author} {\bibinfo {author} {\bibfnamefont {S.}~\bibnamefont {Kundu}}\ and\ \bibinfo {author} {\bibfnamefont {N.}~\bibnamefont {Makri}},\ }\bibfield  {title} {\enquote {\bibinfo {title} {Pathsum: A c++ and fortran suite of fully quantum mechanical real-time path integral methods for (multi-) system+ bath dynamics},}\ }\href@noop {} {\bibfield  {journal} {\bibinfo  {journal} {J. Chem. Phys.}\ }\textbf {\bibinfo {volume} {158}},\ \bibinfo {pages} {224801} (\bibinfo {year} {2023})}\BibitemShut {NoStop}%
\bibitem [{\citenamefont {Bose}\ and\ \citenamefont {Walters}(2022)}]{bose2022multisite}%
  \BibitemOpen
  \bibfield  {author} {\bibinfo {author} {\bibfnamefont {A.}~\bibnamefont {Bose}}\ and\ \bibinfo {author} {\bibfnamefont {P.~L.}\ \bibnamefont {Walters}},\ }\bibfield  {title} {\enquote {\bibinfo {title} {A multisite decomposition of the tensor network path integrals},}\ }\href@noop {} {\bibfield  {journal} {\bibinfo  {journal} {J. Chem. Phys.}\ }\textbf {\bibinfo {volume} {156}},\ \bibinfo {pages} {024101} (\bibinfo {year} {2022})}\BibitemShut {NoStop}%
\bibitem [{\citenamefont {Hubig}, \citenamefont {McCulloch},\ and\ \citenamefont {Schollw{\"o}ck}(2017)}]{hubig2017generic}%
  \BibitemOpen
  \bibfield  {author} {\bibinfo {author} {\bibfnamefont {C.}~\bibnamefont {Hubig}}, \bibinfo {author} {\bibfnamefont {I.}~\bibnamefont {McCulloch}}, \ and\ \bibinfo {author} {\bibfnamefont {U.}~\bibnamefont {Schollw{\"o}ck}},\ }\bibfield  {title} {\enquote {\bibinfo {title} {Generic construction of efficient matrix product operators},}\ }\href@noop {} {\bibfield  {journal} {\bibinfo  {journal} {Phys. Rev. B}\ }\textbf {\bibinfo {volume} {95}},\ \bibinfo {pages} {035129} (\bibinfo {year} {2017})}\BibitemShut {NoStop}%
\bibitem [{\citenamefont {Ren}\ \emph {et~al.}(2020)\citenamefont {Ren}, \citenamefont {Li}, \citenamefont {Jiang},\ and\ \citenamefont {Shuai}}]{ren2020general}%
  \BibitemOpen
  \bibfield  {author} {\bibinfo {author} {\bibfnamefont {J.}~\bibnamefont {Ren}}, \bibinfo {author} {\bibfnamefont {W.}~\bibnamefont {Li}}, \bibinfo {author} {\bibfnamefont {T.}~\bibnamefont {Jiang}}, \ and\ \bibinfo {author} {\bibfnamefont {Z.}~\bibnamefont {Shuai}},\ }\bibfield  {title} {\enquote {\bibinfo {title} {A general automatic method for optimal construction of matrix product operators using bipartite graph theory},}\ }\href@noop {} {\bibfield  {journal} {\bibinfo  {journal} {J. Chem. Phys.}\ }\textbf {\bibinfo {volume} {153}} (\bibinfo {year} {2020})}\BibitemShut {NoStop}%
\bibitem [{\citenamefont {{\c{C}}ak{\i}r}, \citenamefont {Milbradt},\ and\ \citenamefont {Mendl}(2025)}]{ccakir2025optimal}%
  \BibitemOpen
  \bibfield  {author} {\bibinfo {author} {\bibfnamefont {H.}~\bibnamefont {{\c{C}}ak{\i}r}}, \bibinfo {author} {\bibfnamefont {R.~M.}\ \bibnamefont {Milbradt}}, \ and\ \bibinfo {author} {\bibfnamefont {C.~B.}\ \bibnamefont {Mendl}},\ }\bibfield  {title} {\enquote {\bibinfo {title} {Optimal symbolic construction of matrix product operators and tree tensor network operators},}\ }\href@noop {} {\bibfield  {journal} {\bibinfo  {journal} {Phys. Rev. B}\ }\textbf {\bibinfo {volume} {112}},\ \bibinfo {pages} {035101} (\bibinfo {year} {2025})}\BibitemShut {NoStop}%
\bibitem [{\citenamefont {Haegeman}\ \emph {et~al.}(2016)\citenamefont {Haegeman}, \citenamefont {Lubich}, \citenamefont {Oseledets}, \citenamefont {Vandereycken},\ and\ \citenamefont {Verstraete}}]{haegeman2016unifying}%
  \BibitemOpen
  \bibfield  {author} {\bibinfo {author} {\bibfnamefont {J.}~\bibnamefont {Haegeman}}, \bibinfo {author} {\bibfnamefont {C.}~\bibnamefont {Lubich}}, \bibinfo {author} {\bibfnamefont {I.}~\bibnamefont {Oseledets}}, \bibinfo {author} {\bibfnamefont {B.}~\bibnamefont {Vandereycken}}, \ and\ \bibinfo {author} {\bibfnamefont {F.}~\bibnamefont {Verstraete}},\ }\bibfield  {title} {\enquote {\bibinfo {title} {Unifying time evolution and optimization with matrix product states},}\ }\href@noop {} {\bibfield  {journal} {\bibinfo  {journal} {Phys. Rev. B}\ }\textbf {\bibinfo {volume} {94}},\ \bibinfo {pages} {165116} (\bibinfo {year} {2016})}\BibitemShut {NoStop}%
\bibitem [{\citenamefont {Li}, \citenamefont {Ren},\ and\ \citenamefont {Shuai}(2020)}]{li2020numerical}%
  \BibitemOpen
  \bibfield  {author} {\bibinfo {author} {\bibfnamefont {W.}~\bibnamefont {Li}}, \bibinfo {author} {\bibfnamefont {J.}~\bibnamefont {Ren}}, \ and\ \bibinfo {author} {\bibfnamefont {Z.}~\bibnamefont {Shuai}},\ }\bibfield  {title} {\enquote {\bibinfo {title} {Numerical assessment for accuracy and gpu acceleration of td-dmrg time evolution schemes},}\ }\href@noop {} {\bibfield  {journal} {\bibinfo  {journal} {J. Chem. Phys.}\ }\textbf {\bibinfo {volume} {152}},\ \bibinfo {pages} {024127} (\bibinfo {year} {2020})}\BibitemShut {NoStop}%
\bibitem [{\citenamefont {Shao}\ and\ \citenamefont {Makri}(2001)}]{shao2001iterative}%
  \BibitemOpen
  \bibfield  {author} {\bibinfo {author} {\bibfnamefont {J.}~\bibnamefont {Shao}}\ and\ \bibinfo {author} {\bibfnamefont {N.}~\bibnamefont {Makri}},\ }\bibfield  {title} {\enquote {\bibinfo {title} {Iterative path integral calculation of quantum correlation functions for dissipative systems},}\ }\href@noop {} {\bibfield  {journal} {\bibinfo  {journal} {Chem. Phys.}\ }\textbf {\bibinfo {volume} {268}},\ \bibinfo {pages} {1--10} (\bibinfo {year} {2001})}\BibitemShut {NoStop}%
\bibitem [{\citenamefont {Shao}\ and\ \citenamefont {Makri}(2002)}]{shao2002iterative}%
  \BibitemOpen
  \bibfield  {author} {\bibinfo {author} {\bibfnamefont {J.}~\bibnamefont {Shao}}\ and\ \bibinfo {author} {\bibfnamefont {N.}~\bibnamefont {Makri}},\ }\bibfield  {title} {\enquote {\bibinfo {title} {Iterative path integral formulation of equilibrium correlation functions for quantum dissipative systems},}\ }\href@noop {} {\bibfield  {journal} {\bibinfo  {journal} {J. Chem. Phys.}\ }\textbf {\bibinfo {volume} {116}},\ \bibinfo {pages} {507--514} (\bibinfo {year} {2002})}\BibitemShut {NoStop}%
\bibitem [{\citenamefont {Feynman}\ and\ \citenamefont {Vernon~Jr}(2000)}]{feynman2000theory}%
  \BibitemOpen
  \bibfield  {author} {\bibinfo {author} {\bibfnamefont {R.~P.}\ \bibnamefont {Feynman}}\ and\ \bibinfo {author} {\bibfnamefont {F.}~\bibnamefont {Vernon~Jr}},\ }\bibfield  {title} {\enquote {\bibinfo {title} {The theory of a general quantum system interacting with a linear dissipative system},}\ }\href@noop {} {\bibfield  {journal} {\bibinfo  {journal} {Ann. Phys.}\ }\textbf {\bibinfo {volume} {281}},\ \bibinfo {pages} {547--607} (\bibinfo {year} {2000})}\BibitemShut {NoStop}%
\bibitem [{\citenamefont {Leggett}\ \emph {et~al.}(1987)\citenamefont {Leggett}, \citenamefont {Chakravarty}, \citenamefont {Dorsey}, \citenamefont {Fisher}, \citenamefont {Garg},\ and\ \citenamefont {Zwerger}}]{leggett1987dynamics}%
  \BibitemOpen
  \bibfield  {author} {\bibinfo {author} {\bibfnamefont {A.~J.}\ \bibnamefont {Leggett}}, \bibinfo {author} {\bibfnamefont {S.}~\bibnamefont {Chakravarty}}, \bibinfo {author} {\bibfnamefont {A.~T.}\ \bibnamefont {Dorsey}}, \bibinfo {author} {\bibfnamefont {M.~P.}\ \bibnamefont {Fisher}}, \bibinfo {author} {\bibfnamefont {A.}~\bibnamefont {Garg}}, \ and\ \bibinfo {author} {\bibfnamefont {W.}~\bibnamefont {Zwerger}},\ }\bibfield  {title} {\enquote {\bibinfo {title} {Dynamics of the dissipative two-state system},}\ }\href@noop {} {\bibfield  {journal} {\bibinfo  {journal} {Rev. Mod. Phys.}\ }\textbf {\bibinfo {volume} {59}},\ \bibinfo {pages} {1} (\bibinfo {year} {1987})}\BibitemShut {NoStop}%
\bibitem [{\citenamefont {Guo}\ and\ \citenamefont {Chen}(2024)}]{guo2024efficient}%
  \BibitemOpen
  \bibfield  {author} {\bibinfo {author} {\bibfnamefont {C.}~\bibnamefont {Guo}}\ and\ \bibinfo {author} {\bibfnamefont {R.}~\bibnamefont {Chen}},\ }\bibfield  {title} {\enquote {\bibinfo {title} {Efficient construction of the feynman-vernon influence functional as matrix product states},}\ }\href@noop {} {\bibfield  {journal} {\bibinfo  {journal} {SciPost Physics Core}\ }\textbf {\bibinfo {volume} {7}},\ \bibinfo {pages} {063} (\bibinfo {year} {2024})}\BibitemShut {NoStop}%
\bibitem [{\citenamefont {Ishizaki}\ and\ \citenamefont {Fleming}(2009)}]{fmo_sys}%
  \BibitemOpen
  \bibfield  {author} {\bibinfo {author} {\bibfnamefont {A.}~\bibnamefont {Ishizaki}}\ and\ \bibinfo {author} {\bibfnamefont {G.~R.}\ \bibnamefont {Fleming}},\ }\bibfield  {title} {\enquote {\bibinfo {title} {Theoretical examination of quantum coherence in a photosynthetic system at physiological temperature},}\ }\href@noop {} {\bibfield  {journal} {\bibinfo  {journal} {Proc. Natl. Acad. Sci.}\ }\textbf {\bibinfo {volume} {106}},\ \bibinfo {pages} {17255--17260} (\bibinfo {year} {2009})}\BibitemShut {NoStop}%
\bibitem [{\citenamefont {Adolphs}\ and\ \citenamefont {Renger}(2006)}]{fmo_sys2}%
  \BibitemOpen
  \bibfield  {author} {\bibinfo {author} {\bibfnamefont {J.}~\bibnamefont {Adolphs}}\ and\ \bibinfo {author} {\bibfnamefont {T.}~\bibnamefont {Renger}},\ }\bibfield  {title} {\enquote {\bibinfo {title} {How proteins trigger excitation energy transfer in the fmo complex of green sulfur bacteria},}\ }\href@noop {} {\bibfield  {journal} {\bibinfo  {journal} {Biophys. J.}\ }\textbf {\bibinfo {volume} {91}},\ \bibinfo {pages} {2778--2797} (\bibinfo {year} {2006})}\BibitemShut {NoStop}%
\bibitem [{\citenamefont {Ambrosek}\ \emph {et~al.}(2012)\citenamefont {Ambrosek}, \citenamefont {Kohn}, \citenamefont {Schulze},\ and\ \citenamefont {Kühn}}]{ambrosek2012quantum}%
  \BibitemOpen
  \bibfield  {author} {\bibinfo {author} {\bibfnamefont {D.}~\bibnamefont {Ambrosek}}, \bibinfo {author} {\bibfnamefont {A.}~\bibnamefont {Kohn}}, \bibinfo {author} {\bibfnamefont {J.}~\bibnamefont {Schulze}}, \ and\ \bibinfo {author} {\bibfnamefont {O.}~\bibnamefont {Kühn}},\ }\bibfield  {title} {\enquote {\bibinfo {title} {Quantum chemical parametrization and spectroscopic characterization of the frenkel exciton hamiltonian for a j-aggregate forming perylene bisimide dye},}\ }\href@noop {} {\bibfield  {journal} {\bibinfo  {journal} {J. Phys. Chem. A}\ }\textbf {\bibinfo {volume} {116}},\ \bibinfo {pages} {11451--11458} (\bibinfo {year} {2012})}\BibitemShut {NoStop}%
\bibitem [{\citenamefont {Kundu}\ and\ \citenamefont {Makri}(2020)}]{kundu2020modular}%
  \BibitemOpen
  \bibfield  {author} {\bibinfo {author} {\bibfnamefont {S.}~\bibnamefont {Kundu}}\ and\ \bibinfo {author} {\bibfnamefont {N.}~\bibnamefont {Makri}},\ }\bibfield  {title} {\enquote {\bibinfo {title} {Modular path integral for finite-temperature dynamics of extended systems with intramolecular vibrations},}\ }\href@noop {} {\bibfield  {journal} {\bibinfo  {journal} {J. Chem. Phys.}\ }\textbf {\bibinfo {volume} {153}},\ \bibinfo {pages} {044124} (\bibinfo {year} {2020})}\BibitemShut {NoStop}%
\bibitem [{\citenamefont {Kundu}\ and\ \citenamefont {Makri}(2021)}]{kundu2021efficient}%
  \BibitemOpen
  \bibfield  {author} {\bibinfo {author} {\bibfnamefont {S.}~\bibnamefont {Kundu}}\ and\ \bibinfo {author} {\bibfnamefont {N.}~\bibnamefont {Makri}},\ }\bibfield  {title} {\enquote {\bibinfo {title} {Efficient matrix factorisation of the modular path integral for extended systems},}\ }\href@noop {} {\bibfield  {journal} {\bibinfo  {journal} {Mol. Phys.}\ }\textbf {\bibinfo {volume} {119}},\ \bibinfo {pages} {e1797200} (\bibinfo {year} {2021})}\BibitemShut {NoStop}%
\end{thebibliography}%

\end{document}